\documentstyle[aps,epsfig]{revtex}

        \newcommand{\fett}[1]{\mbox{\boldmath$#1$\unboldmath}}
        
        \newcommand{\Dirac}{\mbox{$\not\!\!D$}}
        \newcommand{\be}{\begin{equation}}
        \newcommand{\ee}{\end{equation}}
        \newcommand{\bea}{\begin{eqnarray}}
        \newcommand{\eea}{\end{eqnarray}}
      
\begin{document}   

\date{\today}
\begin{center}
{\Large\bf Mesoscopic QCD and the ${\fett \theta}$-Vacua}
\\[1cm]
Jonathan T.\ Lenaghan\footnote{Email: lenaghan@nbi.dk} and Thomas Wilke\footnote{Email: wilke@nbi.dk}
\\ ~~ \\
{\it The Niels Bohr Institute, Blegdamsvej 17, DK-2100} \\
{\it Copenhagen \O, Denmark}
\\ ~~ \\ ~~ \\
\end{center}

\begin{abstract}
The partition function of QCD is analyzed for an arbitrary number of
flavors, $N_f$, and arbitrary 
quark masses including the contributions from all
topological sectors in the Leutwyler--Smilga regime.  For given $N_f$
and arbitrary vacuum angle, $\theta$, the partition function can be
reduced to $N_f-2$ angular integrations of single Bessel functions.
For two and three flavors, the $\theta$ dependence of the QCD vacuum
is studied in detail.  For
$N_f=$ 2 and 3, the chiral condensate decreases monotonically as
$\theta$ increases from zero to $\pi$ and the chiral condensate
develops a cusp at $\theta=\pi$ for degenerate quark masses in the
macroscopic limit.  We find a discontinuity at $\theta=\pi$ in the first
derivative of the energy density with respect to $\theta$ for
degenerate quark masses.  This corresponds to the first--order phase
transition in which CP is spontaneously broken, known as Dashen's 
phenomena.
\end{abstract}

\section{Introduction}

For $N_f$ flavors of massless quarks, the Lagrangian of quantum
chromodynamics (QCD) is invariant under the global symmetry group
$SU(N_f)_{R} \times SU(N_f)_{L} \times U(1)_{A}$ at the classical
level.  At the quantum level, however, the $SU(N_f)_{R} \times
SU(N_f)_{L}$ symmetry is spontaneously broken to the diagonal subgroup
of vector transformations, $SU(N_f)_{R+L}=SU(N_f)_{V}$, by an
expectation value for the quark-antiquark condensate.  The $U(1)_{A}$
symmetry is explicitly broken to $Z(N_f)_A$ by a non-vanishing
topological susceptibility \cite{thooft}.  For scales well below
$\Lambda$, the typical hadronic mass scale, the effects of the
explicit breaking of the $U(1)_A$ symmetry can be essentially ignored
and the dynamics of the theory are dominated by the $N_f^2-1$
Goldstone bosons that arise from the spontaneous breaking of the
chiral symmetry.  For small, nonzero quark masses, these excitations
become pseudo--Goldstone bosons.

In addition to the quark masses, the QCD Lagrangian depends on another
parameter, $\theta$, the so-called vacuum angle.  Though this
parameter explicitly violates the discrete CP symmetry for all
non-integer multiples of $\pi$, experimental results for the value of
the neutron dipole moment constrain $\theta$ to be zero within a
deviation of less than $10^{-9}$ \cite{Smith:1990ke,Altarev:1992cf}.
The reason why $\theta=0$, however, is still poorly understood, and
there is considerable theoretical interest in the physics of QCD at
$\theta \neq 0$.  The most well-known example is the Veneziano--Witten
formula which in the large-$N_c$ limit relates the mass of the $\eta'$
meson to the second derivative of the pure Yang--Mills vacuum energy
with respect to $\theta$ \cite{Witten1,Veneziano1}.  Another
phenomenologically motivated example is the work of Refs.\
\cite{Kharzeev:1998kz,Buckley:2000mv,Ahrensmeier:2000pg,Buckley:2001aa}
in which it is argued that metastable states in which $\theta$ is
effectively nonzero may be created in ultrarelativistic heavy ion
collisions.

Physics at $\theta \neq 0$ is inherently nonperturbative and one must
rely either on effective theories or perform lattice gauge theory
simulations \cite{Schierholz:1995qg}.  The latter approach is
prohibitively difficult at present because at nonzero values of
$\theta$ the action is complex and standard importance sampling
methods that are usually employed are no longer applicable.  This
difficulty is similar to that for nonzero values of the baryonic
chemical potential, but there has been recent progress in solving the
complex action problem in simpler models
\cite{Chandrasekharan:1999cm,Alford:2001ug}.  While many observables
associated with the quark masses have been studied extensively in
lattice gauge theory computations, properties of the QCD vacuum
associated with $\theta$ such as the topological susceptibility have
only recently been given extensive attention by the lattice community
\cite{Hart:2000hy,Alles:2000cg,Durr:2001ty,Bali:2001gk,Bertle:2001xd,Hasenfratz:2001wd,AliKhan:2001ym,Hart:2001pj}.
So, most of our knowledge of the physics of QCD at $\theta \neq 0$ has
been gleaned from effective theories.  One of the most striking
examples of novel physics is the emergence of two CP violating
degenerate vacua separated by a potential energy barrier at
$\theta=\pi$ known as Dashen's phenomenon \cite{Dashen}.  While QCD is
seemingly invariant under the discrete CP group for $\theta$ equal to
integer values of $\pi$, this invariance is spontaneously broken in a
first--order phase transition at $\theta=\pi$.  Chiral effective
Lagrangians in the large-$N_c$ limit were used to investigate this and
other properties of the $\theta$-vacua in Refs.\ \cite{diVecchia} and
\cite{Witten:1980sp}.  Subsequent elaborations and refinements using
this approach were made in Refs.\
\cite{Creutz:1995wf,Evans:1997eq,Halperin:1998rc,Fugleberg:1999kk,Smilga:1999dh,Tytgat:1999yx}.
Additionally, random matrix models \cite{Janik:2001hk,Janik:1999bj} as
well as numerical simulations of $CP^{N-1}$ models
\cite{Plefka:1997ks,Plefka:1997tz,Yoneyama:1999sj,Burkhalter:2001hu}
have also been employed to investigate the physics of nonzero values
of $\theta$.

In this work, we investigate the properties of QCD at $\theta \neq 0$
in the Leutwyler--Smilga finite volume regime \cite{Leutwyler:1992yt},
also known as the mesoscopic regime.  Here, chiral perturbation theory
is valid but the volume is taken such that the Goldstone modes are
constant and their kinetic term can be ignored.  This approach differs
from those of Refs.\
\cite{diVecchia,Witten:1980sp,Creutz:1995wf,Evans:1997eq,Halperin:1998rc,Fugleberg:1999kk,Smilga:1999dh,Tytgat:1999yx}, 
in which the large-$N_c$ is taken and the Lagrangian is studied
to lowest order in the chiral fields.  The Leutwyler--Smilga 
regime, on the
other hand, enables exact, analytical predictions to be made.  Indeed,
this limit is realized in many lattice simulations
\cite{Verbaarschot:2000dy}.  Many of the low-energy aspects of QCD
have been studied using effective chiral Lagrangians in a finite
volume \cite{Gasser:1987ah}.  In Ref.\ \cite{Leutwyler:1992yt}, Leutwyler and
Smilga demonstrated that such theories contain information beyond pion
dynamics and the spontaneous breaking of chiral symmetry, including
the low-lying spectrum of the QCD Dirac operator and the topology of
the gauge fields.  The purpose of this work is twofold.  First, we
reduce the QCD partition function for $N_f\ge3$ in the
Leutwyler-Smilga scaling regime including the contributions from all
topological sectors to $N_f-2$ angular integrations over single Bessel
functions.  The latter half of this work is devoted to examining the
vacuum properties of QCD in the Leutwyler-Smilga scaling regime at
nonzero values of $\theta$.

The partition functions for one and two flavors of quarks including
contributions from all topological sectors were computed in Refs.\
\cite{Jackson:1996jb} and \cite{Damgaard:1998tr}.  As emphasized in
Ref.\ \cite{Damgaard1}, the contributions to the partition function
from sectors with nonzero topological charge are crucial in the finite
volume regime.  For example, if one considers the partition function
at any fixed topological charge (with the exception of $\nu=0$), then
chiral condensate diverges as one of the quark masses approaches zero.
This completely unphysical behavior is remedied after summing over all
topological charges, a procedure which becomes significantly more
complicated for three or more quark flavors.  We show in the following
that the summation over all topological sectors can be performed for
arbitrary $N_f$ and $\theta$ with the reduction of the partition
function to $N_f-2$ angular integrations over single Bessel functions.

This paper is organized as follows.  In Sec.\ \ref{sec:partition}, we
briefly review the reduction of the QCD partition function to a finite
dimensional group integral in the Leutwyler-Smilga scaling regime.  In
Sec.\ \ref{sec:sum}, we review the summation over all topological charges
for the partition function for $N_f=$1 and 2. We then calculate the
summation for arbitrary $N_f$ and quark masses.  While the full
partition function for $N_f=2$ has been known for some time
\cite{Leutwyler:1992yt}, there has not been a detailed study of the
vacuum properties of this theory for $\theta\neq0$.  In Sec.\
\ref{sec:twoflavors}, we examine the $\theta$ dependence of the chiral
condensate, chiral susceptibility, topological density and topological
susceptibility for two flavors.  In Sec.\ \ref{sec:threeflavors}, we
extend this discussion to three flavors.  Finally, we give concluding
remarks in Sec.\ \ref{sec:theend}.

\section{Partition Functions in the Leutwyler--Smilga Regime} 
\label{sec:partition}

Consider a $SU(N_c)$ gauge theory with $N_f$ flavors of fermions in the
fundamental representation on a four-dimensional
torus of volume $V=L^4$.  The Lagrangian is  
\be \label{eq:Lqcd}
{\cal L} = \frac{1}{4g^2}\, F_{\mu \nu}^a F_{\mu \nu}^a - \sum_{f=1}^{N_f}
        {\bar \psi}_{f} \left(i\Dirac  - m_f \right)
        \psi_{f}  - i\, \theta\,  F_{\mu \nu}^a \tilde{F}^{a}_{\mu \nu} \,\, ,
\ee 
where $\tilde{F}^a_{\mu \nu} = \frac{1}{2} \epsilon_{\mu \nu \alpha
\beta} F_{\alpha \beta}^a$.  The last term in eq.\ (\ref{eq:Lqcd}) 
is a total derivative and so does not affect the field equations or
any of the perturbative aspects of the theory.  The integral 
of this term over the four-volume is quantized and is given by
\be
\nu = \frac{1}{32 \pi^2} \int d^4x \, 
        \epsilon_{\mu\nu\rho\sigma}
        F_{\mu\nu}^a(x) F_{\rho\sigma}^a(x) \in {\bf Z} \,\, .
\label{eq202}
\ee
This quantity, known as the topological charge, 
contributes a phase factor to the path integral and is associated 
with transitions between topologically nontrivial gauge field 
configurations.
The full partition function is given 
by a weighted sum over the partition function for each 
topological sector:
\be\label{eq204}
{\cal Z}^{(N_f)}(\theta,\{m_{k}\}) = \sum_{\nu=-\infty}^{\infty} 
e^{i \nu \theta} {\cal Z}^{(N_f)}_{\nu}(\{m_{k}\}) \,\, .
\ee
The individual contributions are 
\be
{\cal Z}^{(N_f)}_{\nu}(\{m_{k}\}) = 
\int \left[ dA \right]_{\nu} \prod_{f=1}^{N_f} 
\det\left[i\Dirac[A] - m_f \right] 
\exp\left(-\int_Vd^4x{\cal L}_{YM}[A]\right) \,\, ,
\label{eq203}
\ee
where the integration is taken only over gauge field configurations with 
topological charge $\nu$.

In principle, all of the observables of QCD can be calculated by
evaluating eq.\ (\ref{eq204}) and its various functional derivatives.
Typically, however, this can only be done in some approximation since
the integration is taken over an infinite dimensional functional space
and there are ultraviolet and infrared divergences.  Formulating the
theory on a discrete, Euclidean lattice has the advantage of
eliminating the ultraviolet divergences on account of a nonzero
lattice spacing, $a$, and the infrared divergences on account of a
necessarily finite volume.  A second approach is to analyze the
partition function for the effective theory which describes the
low--energy behavior.  For QCD, this is chiral perturbation theory.
Restricting the Euclidean four-volume,
$V=L^4$, to the range
\be \label{eq:range} 
\frac{1}{\Lambda} \ll L \ll \frac{1}{m_\pi} \,\, ,  
\ee
where $\Lambda$ is the typical hadronic scale and $m_\pi$ is the mass
of the Goldstone excitations, results in tremendous simplifications
and {\em exact} analytical predictions are possible
\cite{Gasser:1987ah,Leutwyler:1992yt}.  
This is possible since the lower limit
ensures that the partition function is dominated by Goldstone modes
and the upper limit ensures that these modes are constant, i.e. that the
kinetic term in the partition function factorizes from the mass 
dependent term.

There have been a number of advances in recent years towards 
evaluating the QCD finite volume partition function.  With the
four-volume taken according to eq.\ (\ref{eq:range}), the partition
function was shown in 
Refs.\ \cite{Gasser:1987ah,Leutwyler:1992yt} to reduce to
the finite dimensional group integration
\be 
{\cal Z}^{(N_f)}(\theta,\{m_i\},V) = \int_{SU(N_f)} \, dU \, \exp\left[V \, \Sigma 
        \, {\rm Re} \left( e^{i\theta / N_f} {\rm Tr} 
        {\cal M} U^{\dagger} \right) \right] \,\, ,
\label{eq31}
\ee
where ${\cal M}={\rm diag}(m_1,\ldots,m_{N_f})$ is the mass matrix for
the quark fields and $\Sigma$ is the chiral condensate in the chiral
and infinite volume limit.  We shall henceforth refer to $\Sigma$ as
the macroscopic chiral condensate.  The integration is taken over the
group manifold of the Goldstone modes, $SU(N_f)$.  Note that the
dependence on the volume, the macroscopic chiral condensate and the
quark masses is only through the dimensionless scaling variable
\be
\mu_i=\Sigma\, V\, m_i.
\ee

The Fourier coefficients conjugate to ${\cal Z}^{(N_f)}(\theta,\{\mu_i\})$
are obtained by taking the Fourier transformation of eq.\ (\ref{eq204}),
\be
{\cal Z}_{\nu}^{(N_f)}(\{\mu_{i}\})
=\frac{1}{2 \pi}\int\limits_0^{2\pi}d\theta\,
{\cal Z}^{(N_f)}(\theta,\{\mu_i\})e^{-i\nu\theta} \,\,.
\ee
They have been computed for arbitrary $N_f$ and $N_c \ge 3$
\cite{Brower:1981vt,Jackson:1996jb}, and can be expressed as 
\be
{\cal Z}_{\nu}^{(N_f)}(\{\mu_{i}\},V) = 
\frac{\det {\cal A_\nu}(\{\mu_{i}\})}{\Delta(\{\mu_{i}^2\})} \,\, , 
\label{eq:Zdef}
\ee
where 
\begin{eqnarray*}
{\cal A_\nu}(\{\mu_{i}\}) = 
\left( \begin{array}{cccc} 
I_\nu(\mu_1) & \mu_1 I_{\nu+1}(\mu_1) & 
\cdots & \mu_1^{N_f-1} I_{\nu+N_f-1}(\mu_1) \\
\vdots & \vdots & \ddots & \vdots \\
I_\nu(\mu_{N_f}) & \mu_{N_f} I_{\nu+1}(\mu_{N_f}) & \cdots & 
\mu_{N_f}^{N_f-1} I_{\nu+N_f-1}(\mu_{N_f})
\end{array}\right)
\end{eqnarray*}
is a $N_f \times N_f$ matrix and $I_n$ is a modified Bessel 
function of order $n$.  The denominator, 
\be 
\Delta(\{\mu_{i}^2\}) = \prod_{i>j} (\mu_{i}^{2} - \mu_{j}^{2}) \,\, ,
\ee
is the Vandermonde determinant.  For equal masses, $\mu_i=\mu$, ${\cal
Z}_{\nu}(\mu)$ can be further simplified using properties of 
Bessel functions and determinants to
give
\be
{\cal Z}_{\nu}^{(N_f)}(\mu) = \left| 
        \begin{array}{cccc}
         I_{\nu}(\mu) & I_{\nu+1}(\mu) & \cdots & \,\,\, I_{\nu+N_f-1}(\mu) \\
         I_{\nu-1}(\mu) & I_{\nu}(\mu) & \cdots & \,\,\, I_{\nu+N_f-2}(\mu) \\
         \vdots & \vdots & \ddots & \vdots \\
         I_{\nu-N_f+1}(\mu) & I_{\nu-N_f+2}(\mu) & \cdots & \,\,\, I_{\nu}(\mu)
        \end{array} 
        \right| \,\, .
\label{eq35}
\ee
Combinatorial formulas for generating the character expansions of the $U(N)$
group were recently given in Ref.\ \cite{Balantekin:2000vn} and used 
to very efficiently derive eq.\ (\ref{eq:Zdef}). 
The close connection to random matrix theory was firmly established 
in Refs.\ \cite{Osborn:1999qb,Damgaard:1999xy,Damgaard:2000ic}.

The vacuum properties of the theory are defined 
by eq.\ (\ref{eq31}) and its derivatives.  
Recall that the first derivative of the logarithm of the partition
function with respect to a parameter determines the mean value of the
conjugate variable to the parameter, while the second derivative
serves as a measure for the fluctuations around the average value.
We focus on 
the derivatives with respect to the quark masses, $m_i$, and 
the vacuum angle, $\theta$.
The chiral condensate for the $i$-th quark species is 
\be 
\Sigma_{(i)}^{(N_f)}(\theta,\{m_{k}\},V) = \frac{1}{VN_f} 
        \frac{\partial}{\partial m_{i}} 
        \log {\cal Z}^{(N_f)}(\theta,\{m_{k}\},V) \,\, ,
\label{eq206}
\ee
with the average value
\be
\Sigma^{(N_f)}(\theta,\{m_{k}\},V) =
\sum_{i=1}^{N_f}\Sigma_{(i)}^{(N_f)}(\theta,\{m_{k}\},V)\;.
\label{eq206b}
\ee
This quantity should approach macroscopic chiral condensate, 
$\Sigma$, at $\theta=0$ and for large values of the scaling 
variable, $\mu$.  
The chiral (scalar) susceptibility is defined as
\be
\chi_{ij}^{(N_f)}(\theta,\{m_{k}\},V)=
\frac{1}{V}\frac{\partial^2}{\partial m_i\partial m_j} 
\log {\cal Z}^{(N_f)}(\theta,\{m_{k}\},V)\;.
\label{eq206c}
\ee

Vacuum properties associated with the vacuum angle, $\theta$, 
can be analogously defined.
The topological density is 
\be 
\sigma(\theta,\{m_{k}\},V) = 
-\frac{1}{V}\frac{\partial}{\partial \theta} 
\log {\cal Z}^{(N_f)}(\theta,\{m_{k}\},V) \,\, ,
\label{eq207}
\ee
which at $\theta=0$ has the interpretation of the mean topological charge.
From eqs.\ (\ref{eq204}) and (\ref{eq:Zdef}), one can see that the
topological density vanishes if $\theta$ is an integer multiple of
$\pi$.  The
topological susceptibility is defined by
\be 
\chi_{\rm top}(\theta,\{m_{i}\},V) 
=-\frac{1}{V}\frac{\partial^2}{\partial \theta^2} 
\log {\cal Z}^{(N_f)}(\theta,\{m_{i}\},V) \,\, ,
\label{eq209}
\ee
which at $\theta=0$ is the 
mean square deviation of the topological charge and is in general nonzero.

\section{Summation Over Topological Charges} \label{sec:sum}

When studying the vacuum properties, e.g. the chiral condensate, in
the Leutwyler--Smilga regime, the necessity of including the
contributions to the partition function from every topological sector
was established in Ref.\ \cite{Damgaard1}.  For one and two quark
flavors, the summation can be performed using standard Bessel function
identities.  For $N_f\ge3$, however, the summation is more complicated
and to the best of our knowledge has not been performed previously in
the literature.  It is instructive to review the derivation of the
results for one and two flavors before proceeding to the derivation
for $N_f\ge3$.

\subsection{${\fett N_f=1}$}

For $N_f=1$, the summation over all topological sectors is
straightforward to calculate.  Using the definition of the generating
function for modified Bessel functions \cite{Watson},
\be 
\exp\left[\frac{x}{2}(t+t^{-1})\right]
=\sum_{\nu=-\infty}^{\infty} t^\nu I_\nu(x)\,\, ,
\label{eq401}
\ee 
the summation can be performed:
\bea
\label{eq402}
{\cal Z}^{(N_f=1)}(\theta,\mu) &=& \sum_{\nu=-\infty}^{\infty} 
e^{i\nu \theta} I_{\nu}(\mu)
\\ \nonumber 
        & = & \exp\left[ \frac{\mu}{2} 
        \left(e^{i\theta}+e^{-i\theta} \right)\right] \\ \nonumber
        &=& \exp\left[\mu \, \cos \theta\right]\,\, .
\eea
This result was first found in Ref.\ \cite{Leutwyler:1992yt}.

\subsection{${\fett N_f=2}$}

For two flavors, the calculation is more involved but the method
generalizes very naturally to $N_f \ge 3$.  From eq.\ (\ref{eq:Zdef}),
the partition function is
\be
\label{eq403}
{\cal Z}^{(N_f=2)}(\theta,\mu_1,\mu_2) = \frac{1}{\mu_2^2-\mu_1^2} 
        \sum_{\nu=-\infty}^{\infty} \, e^{i\nu \theta} \, 
        \Big[ \mu_2 \, I_{\nu}(\mu_1) \, I_{\nu+1}(\mu_2) - 
        \mu_1 \, I_{\nu}(\mu_2) \, I_{\nu+1}(\mu_1) \Big]  \,\, .
\ee
Since the partition function is symmetric in the two scaling 
variables, we consider only the first term between the 
brackets in eq.\ (\ref{eq403}).  Following Ref.\ \cite{Watson}, 
we use the contour integral representation for $I_{\nu}(x)$ which follows 
from eq.\ (\ref{eq401}),
\be
I_{\nu}(x) = \oint \frac{ds}{2\pi i} \,s^{-\nu-1} \,
 \exp\left[\frac{x}{2} \left(s+s^{-1}\right)\right] \,\, ,
\ee
where the contour is the standard Bessel function contour surrounding
the negative real axis and the origin in a counterclockwise fashion.
Using this representation, the summation can be calculated:
\begin{mathletters}
\bea
\label{eq404}
\sum_{\nu=-\infty}^{\infty} e^{i\nu \theta} I_{\nu}(\mu_1)I_{\nu+1}(\mu_2)
   &=& \sum_{\nu=-\infty}^{\infty}\, e^{i\nu \theta} \, I_{\nu}(\mu_1)
          \oint \frac{ds}{2\pi i} \,s^{-\nu-2} \,
 \exp\left[\frac{\mu_2}{2} \left(s+s^{-1}\right)\right] \\  
        &=& \oint \frac{ds}{2\pi i} \,s^{-2}    \exp\left[\frac{\mu_2}{2} \left(s+s^{-1}\right)\right] \sum_\nu
          \left(e^{i\theta}s^{-1}\right)^\nu I_\nu(\mu_1) \\ 
        &=& \oint \frac{ds}{2\pi i} \, s^{-2} \, 
  \exp\left[\frac{\mu_2}{2} \left(s+s^{-1}\right)\right] \, 
  \exp\left[\frac{\mu_1}{2} \left(s\, e^{-i\theta} + 
          s^{-1}\,e^{i\theta}\right)\right] \,\, .
\eea
\end{mathletters}
By making the change of variables,
\begin{mathletters}
\bea 
\label{eq405a}
\omega &=& \frac{(\mu_2+\mu_1 e^{-i\theta})}{\mu_{12}(\theta)}s \\
\label{eq405b}
\omega^{-1} &=& \frac{(\mu_2+\mu_1e^{i\theta})}{\mu_{12}(\theta)} s^{-1}\; ,
\eea
where 
\be
\label{eq406}
\mu_{12}(\theta) \equiv \sqrt{\mu_1^2+\mu_2^2+2 \, \mu_1 \, \mu_2 \,
\cos\theta}\;,
\ee
\end{mathletters}
eq.\ (\ref{eq404}) can then be simplified to yield
\bea
\label{eq407}
\sum_{\nu=-\infty}^{\infty}\,e^{i\nu \theta}\,I_{\nu}(\mu_1)\,I_{\nu+1}(\mu_2)
        &=& \frac{\mu_2+\mu_1 e^{-i\theta}}{\mu_{12}(\theta)} \, 
          \oint \frac{d\omega}{2\pi i} \, \omega^{-2} 
   \exp\left[\frac{\mu_{12}(\theta)}{2}(\omega + \omega^{-1})\right] 
\\ \nonumber
        &=& \frac{\mu_2+\mu_1 e^{-i\theta}}{\mu_{12}(\theta)} \, 
          I_1(\mu_{12}(\theta)) \,\, .
\eea
Combining the two contributions, the partition function for $N_f=2$ is 
\be 
\label{eq408}
{\cal Z}^{(N_f=2)}(\theta,\mu_1,\mu_2)
=\frac{I_1(\mu_{12}(\theta))}{\mu_{12}(\theta)} 
\,\, . \label{eq:Z2}
\ee
Equation (\ref{eq406}) defines a reduced mass determined 
by a triangle law.  The triangle has sides $\mu_1$, $\mu_2$
and $\mu_{21}$ and the angle subtended by $\mu_1$ and $\mu_2$ is
$\pi-\theta$.

\subsection{${\fett N_f\ge3}$}

For $N_f\ge3$, we consider cases of degenerate and nondegenerate quark
masses separately.  In the case of nondegenerate quark masses, the
summation over topological charge is a generalization of the summation
for $N_f=2$.  In the process of the derivation, we find interesting
relationships between sums of products of arbitrarily many Bessel
functions and angular integrations over single Bessel functions.

\subsubsection{Nondegenerate quark masses}

From eq.\ (\ref{eq:Zdef}), the quantity that needs
to be calculated is a summation of products of $N_f$ modified Bessel
functions weighted by a phase, since
\be
\label{eq409}
\det {\cal A}_\nu(\{\mu_i\})=\varepsilon_{i_1\ldots i_{N_f}}
\prod_{j=1}^{N_f}\mu_{i_j}^{j-1}I_{\nu+j-1}(\mu_{i_j})\;.
\ee
Each term in the summation is of the form
\bea
\label{eq410}
{\cal B} &=&  \sum_{\nu=-\infty}^{\infty} e^{i\nu \theta} 
I_{\nu+m_1}(x_1)I_{\nu+m_2}(x_2)\cdots I_{\nu+m_{N_f}}(x_{N_f})
\\ \nonumber 
& = & e^{-im_1\theta} \sum_{\nu=-\infty}^{\infty} e^{i\nu \theta} 
I_{\nu}(x_1)I_{\nu+n_2}(x_2)\ldots I_{\nu+n_{N_f}}(x_{N_f}) 
\\ \nonumber
& \equiv & e^{-im_1\theta}\;{\cal C}\;,
\eea
where $n_j=m_j-m_1$.  For the calculation of the partition function, 
${\cal B}$ is actually more general than necessary and the particular 
choices of $x_j=\mu_{i_j}$ and $m_j=j-1$ specialize to 
the appropriate terms in the partition function.  

By expressing all but the first of the modified Bessel functions by 
their contour integral representations, ${\cal C}$ can be written as
\be
\label{eq411}
{\cal C}=\sum_{\nu=-\infty}^{\infty} e^{i\nu \theta} I_{\nu}(x_1)
\prod_{j=2}^{N_f}\oint\left(\frac{ds_j}{2\pi i}s_j^{-\nu-n_j-1}
\exp\left[\frac{x_j}{2}(s_j+s_j^{-1})\right]\right)\;.
\ee
The definition of the generating function for $I_n(x)$,
eq.\ (\ref{eq401}), can be used to perform the summation over $\nu$:
\be
\label{eq412}
{\cal C}=\prod_{j=2}^{N_f}\oint\left(\frac{ds_j}{2\pi i}s_j^{-n_j-1}
\exp\left[\frac{x_j}{2}(s_j+s_j^{-1})\right]\right)
\exp\left[\frac{x_1}{2}\left(\frac{e^{i\theta}}{s_2\ldots s_{N_f}}+
\frac{s_2\ldots s_{N_f}}{e^{i\theta}}\right)\right]\;.
\ee
We now make a change of variables which is suggested by 
eqs.\ (\ref{eq405a}) and (\ref{eq405b}):
\bea
\label{eq413a}
\omega&=&\left(x_2+x_1\frac{s_3\ldots s_{N_f}}{e^{i\theta}}\right)
\frac{s_2}{\psi} \\
\label{eq413b}
\omega^{-1}&=&\left(x_2+x_1\frac{e^{i\theta}}{s_3\ldots s_{N_f}}\right)
\frac{s_2^{-1}}{\psi}\;,
\eea
where 
\be
\label{eq414}
\psi^2=x_1^2+x_2^2+x_1 x_2 \left( \frac{e^{i\theta}}{s_3\ldots s_{N_f}} 
+\frac{s_3\ldots s_{N_f}}{e^{i\theta}} \right)\,\,.
\ee  
The contour integration over $s_2$ is rewritten as 
\bea
\label{eq415}
\lefteqn{\oint \frac{ds_2}{2\pi i} s_2^{-n_2-1} 
\exp\left[\frac{s_2}{2}\left(x_2+x_1\frac{s_3\ldots s_{N_f}}{e^{i\theta}}\right)
+\frac{1}{2s_2}\left(x_2+x_1\frac{e^{i\theta}}{s_3\ldots s_{N_f}}\right)\right]}
\\ \nonumber &=&
\left( \frac{x_2+x_1 s_3\ldots s_{N_f} e^{-i\theta}}{\psi}\right)^{n_2} 
\oint \frac{d\omega}{2\pi i} \, \omega^{-n_2-1} 
        \exp\left(\frac{\psi}{2} (\omega + \omega^{-1})\right)\;.
\\ \nonumber &=&
\left( \frac{x_2+x_1 s_3\ldots s_{N_f} e^{-i\theta}}{\psi}\right)^{n_2} 
I_{n_2}(\psi)
\eea
The expression for ${\cal C}$ is then
\be
\label{eq416}
{\cal C}=
\oint\prod_{j=3}^{N_f}\left(\frac{ds_j}{2\pi i}s_j^{-n_j-1}
\exp\left[\frac{x_j}{2}(s_j+s_j^{-1})\right]\right)
\left( \frac{x_2+x_1 s_3\ldots s_{N_f} e^{-i\theta}}{\psi}\right)^{n_2}
I_{n_2}(\psi)\;.
\ee
Making the second change of variables, $s_k=e^{i\phi_k}$, for
$k=3\ldots,N_f$, and deforming the contours to the unit 
circle, the final expression for ${\cal C}$ is obtained:
\bea
\label{eq417}
{\cal C} &=& \int_{0}^{2\pi} \prod_{j=3}^{N_f}\left(\frac{d\phi_j}{2\pi} \nonumber
\exp[x_j\cos(\phi_j)-in_j\phi_j]\right)
\left(\frac{x_2+x_1 e^{i(\phi_3+\ldots+\phi_{N_f}-\theta)}}
          {\sqrt{x_2^2+x_1^2 + 2 x_2 x_1 \cos(\phi_3+
          \ldots+\phi_{N_f}-\theta)}}\right)^{n_2} \\ \nonumber
        \vspace{1cm} \\ \nonumber
        &\times&
        I_{n_2}\left(\sqrt{x_2^2+x_1^2 + 2 x_2 x_1 \cos(\phi_3+
          \ldots+\phi_{N_f}-\theta)}\right) \,\,. \label{eq:gensum}
\eea
Collecting the various contributions, the partition function 
for $N_f$ quark flavors is 
\be 
\label{eq418}
{\cal Z}^{(N_f)}(\theta,\{\mu_i\}) = 
\frac{1}{\Delta(\{\mu_i^2\})} \varepsilon_{i_1...i_{N_f}}
\left(\prod_{k=1}^{N_f}\mu^{k-1}_{i_k}\right)\mu_{i_2}
S(\mu_{i_1},\mu_{i_2};\mu_{i_3},\dots,\mu_{i_{N_f}}) \,\, ,
\ee
where 
\bea
\label{eq419} \label{eq:S}
\lefteqn{S(\mu_{i_1},\mu_{i_2};\mu_{i_3},\dots,\mu_{i_{N_f}})}
\\ \nonumber &=& 
\int_{0}^{2\pi} \prod_{k=3}^{N_f} \,\left(\frac{d\phi_k}{2\pi} \, 
\exp\left[-i(k-1)\phi_k + \mu_{i_k} \cos(\phi_k)\right] \right) \,
\frac{I_1(\mu_{i_1i_2}(\theta;\phi_3,\ldots,\phi_{N_f}))}
{\mu_{i_1i_2}(\theta;\phi_3,\ldots,\phi_{N_f})} \,\, ,
\eea
and
\be
\label{eq420}
\mu_{i_1i_2}(\theta,\phi_3,\ldots,\phi_{N_f})
= \sqrt{ \mu_{i_1}^2 + \mu_{i_2}^2 + 2 \mu_{i_1} \mu_{i_2}
        \cos\left(\phi_3+\ldots+\phi_{N_f}-\theta\right)} \;.
\ee
Note that $S(\mu_{i_1},\mu_{i_2};\mu_{i_3},\dots,\mu_{i_{N_f}})$ is
symmetric under the interchange of $\mu_{i_1}$ and $\mu_{i_2}$.  

In the case of complete nondegeneracy, $\mu_i\neq \mu_j$ for all
$i\neq j$, the partition function and its derivatives are analytic
functions to all orders.  Any potential nonanalyticities in eq.\
(\ref{eq:S}) and its derivatives occur when
$\mu_{i_1i_2}(\theta;\phi_3,\ldots,\phi_{N_f})$ vanishes which is only
possible if at least two quark masses become degenerate\footnote{This is 
true for finite $\mu_i$.  As shown for $N_f=2$ in Ref.\ \cite{Akemann:2001ir}, there can be 
a nonanalyticity if the two scaling variables are taken to infinity such 
that $\lim_{\mu_1,\mu_2 \to \infty} \frac{(\mu_1-\mu_2)^2}{\mu_1\mu_2} \to 0\displaystyle$
We expect that this limit can be generalized to 
arbitrary $N_f$ in a straightforward manner.}.
  This becomes
apparent when eq.\ (\ref{eq420}) is rewritten in the form 
\bea  \label{eq:cusps}
\lim_{\mu_{i_1},\mu_{i_2}\rightarrow\mu} 
	\mu_{i_1i_2}(\theta,\phi_3,\ldots,\phi_{N_f}) &=& 
	\mu \sqrt{2+2 \cos\left(\phi_3+\ldots+\phi_{N_f}-\theta\right)} \\ \nonumber
	&=& 
	2 \mu \left|\cos\left(\frac{\phi_3+\ldots+\phi_{N_f}-\theta}{2}\right) \right|
\eea
which has a cusp at $\phi_3+\ldots+\phi_{N_f}-\theta=\pi$.  This 
behavior is a necessary condition for the derivatives of eq.\ (\ref{eq418})
to have discontinuities.  On account of the integration, this is 
not a sufficient condition.

\subsubsection{Degenerate Quark Masses}

The limit of equal quark masses, $\mu_i\to\mu$, can be calculated
either by summing over all topological sectors for nondegenerate quark
masses and then taking the limit $\mu_i\to\mu$, or starting from the
equal mass partition function in a given topological sector, eq.\
(\ref{eq35}) and then summing over all topological charges.  Using the
first approach, one can derive equal mass limit from
eqs.~(\ref{eq418})-(\ref{eq420}) and the resulting expression for the
partition function involves derivatives up to order $N_f-1$ of
modified Bessel functions with respect to the masses.  This approach
is quite cumbersome and, moreover, the derivatives are by construction
taken at the cusps given by eq.\ (\ref{eq:cusps}).  The second approach is
more tractable and is the one used here.  After expanding the
determinant in eq.\ (\ref{eq35}),
\be 
{\cal Z}^{(N_f)}_\nu(\mu)=
\varepsilon_{i_1\ldots i_{N_f}}
\prod_{j=1}^{N_f}I_{\nu+j-i_j}(\mu)\,\,,
\label{eq421}
\ee
and using the contour integral representation for each Bessel 
function, the expression for the equal mass partition function 
in a sector of given topological charge becomes
\bea
\label{eq422}
{\cal Z}^{(N_f)}_\nu(\mu) & = &
\oint\prod_{j=1}^{N_f} \left(\frac{ds_j}{2\pi i} 
\exp\left[\frac{\mu}{2}(s_j+s_j^{-1})\right]s_j^{-\nu-j-1}\right)
\varepsilon_{i_1\ldots i_{N_f}} s_1^{i_1}\ldots s_{N_f}^{i_{N_f}}
\\ \nonumber
& = & \int\limits_0^{2\pi}\prod_{j=1}^{N_f} \left(\frac{d\phi_j}{2\pi}
\exp\left[\mu\cos(\phi_j)-i\nu\phi_j-ij\phi_j \right]\right)
\varepsilon_{i_1\ldots i_{N_f}} e^{i \phi_{i_1}}\ldots e^{i N_f \phi_{i_{N_f}}}
\,\, .
\eea
The last equality is obtained by change of variables,
$s_j=e^{i\phi_j}$, deforming the integration contours to the unit
circle, and making use of the identity
\be
\varepsilon_{k_1\ldots k_{N_f}} e^{i \phi_{k_1}}\ldots e^{i N_f \phi_{k_{N_f}}}
=\varepsilon_{k_1\ldots k_{N_f}} e^{i k_1 \phi_{1}}\ldots e^{i k_{N_f} \phi_{N_f}}\,\, .
\ee
The authors of Ref.\ \cite{Leutwyler:1992yt} showed using 
a result of Weyl that 
\be 
{\cal Z}^{(N_f)}_{\nu}(\mu) = \frac{1}{N_{f}!}
        \int_{0}^{2\pi} \prod_{j=1}^{N_f} \left(\frac{d\phi_j}{2\pi}
        \exp\left[\mu\cos(\phi_j)+i\nu\phi_j\right]\right)
        \prod_{k < l} \left|e^{i \phi_k}-e^{i \phi_l}\right|^2 \,\, .
\label{eq423}
\ee
Equations (\ref{eq422}) and (\ref{eq423}) can be transformed 
into one another by observing that
\be
\varepsilon_{i_1\ldots i_{N_f}} e^{i\phi_{i_1}}\ldots 
e^{iN_f\phi_{i_{N_f}}}=\Delta(\{e^{i\phi_j}\})=
\prod_{k < l} \left(e^{i \phi_k}-e^{i \phi_l}\right)
\label{eq424}
\ee
is a Vandermonde determinant, and so 
\be
\prod_{k < l} \left|e^{i \phi_k}-e^{i \phi_l}\right|^2
=\Delta(\{e^{i\phi_j}\})\Delta(\{e^{-i\phi_j}\}) \,\, .
\label{eq425}
\ee
By expanding $\Delta(\{e^{-i\phi_j}\})$, one can transform both
expressions into one another by an appropriate relabeling of 
the integration variables. 

The summation over the topological charges can be performed using
\be
\int_0^{2\pi}\frac{d\phi_1}{2\pi} f(\phi_1,\ldots,\phi_{N_f},\theta)
\sum_{\nu=-\infty}^\infty e^{i\nu(\theta-\phi_1\ldots-\phi_{N_f})}
=
f(\phi_2+\ldots+\phi_{N_f}-\theta,\phi_2,\ldots,\phi_{N_f},\theta)\;,
\label{eq426}
\ee
provided $f$ is $2\pi$-periodic in $\phi_1$.  For the
subsequent integration over $\phi_2$, an integral
representation of the Bessel functions is used. The details of the calculation
are given in the appendix.
The result is 
\bea \label{eq:geneqZ}
{\cal Z}^{(N_f)}(\theta,\mu) &=&  
	\varepsilon_{a_1\ldots a_{N_f}} (a_2-a_1)\prod\limits_{m=3}^{N_f} 
        \left( \int\limits_0^{2\pi} \frac{d\phi_m}{2\pi}
        e^{\mu \cos(\phi_m)}\right) 
	\frac{I_{a_1-a_2}\left(\mu(\phi_3,\ldots,\phi_{N_f},\theta)\right)}
	{\mu(\phi_3,\ldots,\phi_{N_f},\theta)}\\ \nonumber
        &\times&
        \cos\left\{(3-a_1-a_2)(\theta+\phi_3+\ldots
	+\phi_{N_f})/2-(3-a_3)\phi_3-\ldots-
	(N_f-a_{N_f}) \phi_{N_f}\right\}\,\, ,
\eea
where 
\be 
\mu(\phi_3,\ldots,\phi_{N_f},\theta) = 
\mu \sqrt{2+2\cos{\left(\phi_3+\ldots+\phi_{N_f}-\theta\right)}} \,\, .
\ee
The summation over the completely antisymmetric tensor 
can be simplified to give a 
$(N_f-2)$--fold integration over a sum of $N_f-1$ Bessel functions 
multiplied by some phases,
\begin{eqnarray}
{\cal Z}^{(N_f)}(\theta,\mu) &=&
-2e^{-i\theta/2}\int_0^{2\pi}
\prod_{m=3}^{N_f}\left(\frac{d\phi_m}{2\pi}
e^{\mu\cos(\phi_m)-i(m-3/2)\phi_m}\right)
\nonumber\\ &&
\times\sum_{r=1}^{N_f-1}r(-1)^r 
\frac{I_r(\mu(\phi_3,\ldots,\phi_{N_f},\theta))}
{\mu(\phi_3,\ldots,\phi_{N_f},\theta)}
\alpha^{(N_f)}_r(\phi_3,\ldots,\phi_{N_f},\theta)\;.
\end{eqnarray}
The phases $\alpha^{(N_f)}_r(\phi_3,\ldots,\phi_{N_f},\theta)$ are
given by
\begin{equation}
\alpha^{(N_f)}_r(\phi_3,\ldots,\phi_{N_f},\theta)=\sum_{j=1}^{N_f-r}
e^{-i(j+r/2)(\phi_3+\ldots+\phi_{N_f}-\theta)}
\beta^{(N_f)}_{j,j+r}(\phi_3,\ldots,\phi_{N_f})\;,
\end{equation}
with
\begin{equation}
\beta^{(N_f)}_{k,l}(\phi_3,\ldots,\phi_{N_f})=
\left|\begin{array}{ccccccccc}
1 & \cdots & e^{i(k-2)\phi_3} & e^{ik\phi_3} & \cdots 
& e^{i(l-2)\phi_3} & e^{il\phi_3} & \cdots & e^{i(N_f-1)\phi_3} \\
\vdots & & \vdots & \vdots & & \vdots & \vdots & & \vdots  \\
1 & \cdots & e^{i(k-2)\phi_{N_f}} & e^{ik\phi_{N_f}} & \cdots 
& e^{i(l-2)\phi_{N_f}} & e^{il\phi_{N_f}} & \cdots & e^{i(N_f-1)\phi_{N_f}}
\end{array}\right|\;.
\end{equation}
Despite appearances, this last expression is real.  As mentioned
above, the possible emergence of a cusp in the integrand occurs in the
ratios of the Bessel functions and
$\mu(\phi_3,\ldots,\phi_{N_f},\theta)$.  At this stage, it is not {\em
a priori} obvious that there are nonanalyticities after the
integrations are performed.  Thus, one must consider them on a case by
case basis.  This is done for $N_f=2$ and 3 in the following sections.

\section{Vacuum Properties for ${\fett N_f=2}$} \label{sec:twoflavors}

Since the partition function for $N_f=2$, eq.\ (\ref{eq:Z2}), can be
expressed in closed form even after summing over all topological charges, 
calculating the vacuum observables is 
straightforward.  Most of the work in the literature has focused 
upon the behavior of the chiral condensate and the topological 
susceptibility as functions of the scaling variable for either 
$\theta=0$ or in a sector of fixed topological charge.
In this section, we demonstrate that QCD in the Leutwyler--Smilga 
scaling regime exhibits a great deal of nontrivial behavior
at nonzero values of $\theta$ for $N_f=2$.  

\subsection{Partition Function}\label{spectral}

The partition function for two quark flavors is
\be
{\cal Z}^{(N_f=2)}(\theta,\mu_1,\mu_2)=
\frac{I_1\left(\sqrt{\mu_1^2+\mu_2^2+2\mu_1 \mu_2 \cos \theta}\right)}{\sqrt{\mu_1^2+\mu_2^2+2\mu_1 \mu_2 \cos \theta}} \,\, .
\ee
At $\theta=(2n+1) \pi$, where $n$ is an integer, it reduces to 
\be
{\cal Z}^{(N_f=2)}\big(\theta = (2n+1) \pi,\mu_1,\mu_2\big)=
\frac{I_1\left(|\mu_1-\mu_2|\right)}{|\mu_1-\mu_2|} \,\, ,
\ee
i.e. the theory at $\theta=\pi$ is equivalent to the theory at $\theta=0$ but 
taking one quark masses to be negative.  This is in accordance 
with the standard lore.  Taking the two quark masses to be equal, 
however, leads to the surprising result
that the partition function is {\em independent}
of the scaling variable,
\be \label{eq:constpart}
{\cal Z}^{(N_f=2)}\big(\theta=(2n+1)\pi,\mu\big)=\frac{1}{2} \,\, .
\ee
This can be seen also by starting from the equal mass partition function:
\be
{\cal Z}^{(N_f=2)}(\theta,\mu)=
\frac{I_1\left(\mu \, \sqrt{2+2\cos\theta}\right)}{\mu \, \sqrt{2+2\cos\theta}} \,\, .
\ee
Since ${\cal Z}^{(N_f=2)}(\theta,\mu_1,\mu_2)$ is a smooth function of
$\mu_1$, $\mu_2$ and $\theta$, it does not matter in which order the
limits of degenerate masses and $\theta=(2n+1)\pi$ are taken.  This
result may be understood in terms of the microscopic
spectral density.  As shown in Ref.\ \cite{Damgaard1}, after summing
over all topological charges, the full microscopic spectral density is
related to a particular quotient of the partition function for $N_f$
flavors at $\theta$ and the partition function for $N_f+2$ flavors at
$\theta+\pi$.  Thus, the partition function for two flavors at
$\theta=\pi$ is related to the quenched ($N_f=0$) theory at
$\theta=0$.  This fact also demonstrates at a qualitative level that
the partition function becomes independent of the quark masses if and
only if $N_f=2$ and $\theta=\pi$ \cite{poulpriv}. 
It is apparent already at
the level of the partition function that, at least in the
Leutwyler--Smilga scaling regime, QCD is very different qualitatively
at $\theta=n\pi$ than at all other values of $\theta$.

This can also be understood directly from chiral perturbation theory. 
For $N_f=2$, the leading order term in chiral perturbation theory, 
$\Sigma {\rm Re} \left[ {\rm Tr} \left\{ {\cal M} e^{i\theta/N_f} U^{\dagger}
\right\} \right]$, vanishes at $\theta=\pi$ when ${\cal M} = m {\bf 1}$ 
since the trace of an $SU(2)$ matrix is always real.  This term, however, 
is the only relevant term in the Leutwyler--Smilga scaling regime as can 
be seen from eq.\ (\ref{eq31}).  From this observation, it is clear that 
the partition function at $\theta=\pi$ is simply a constant in this case. 
The resolution to this problem was given in Ref.\ \cite{Creutz:1995wf} where 
higher order terms quadratic in masses were included in the effective 
chiral Lagrangian.  A more extensive analysis in terms of chiral 
perturbation theory was given in Refs.\ \cite{Smilga:1999dh,Tytgat:1999yx}.
By including higher order terms in the chiral expansion, the energy 
density then depends once again explicitly on the quark masses. 
These terms, however, are suppressed in the Leutwyler-Smilga regime 
and so we do not include them.

\begin{figure}
\centerline{\epsfig{file=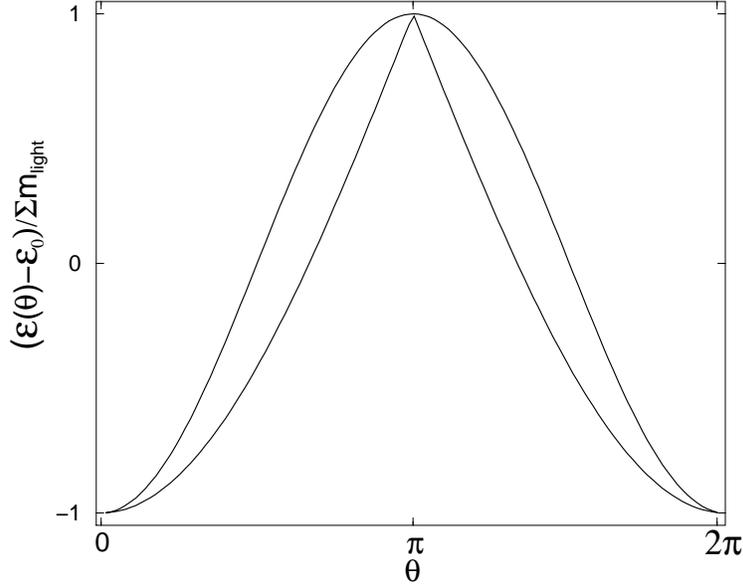,width=10cm}}
\caption{The energy density for $N_f=2$ in the 
macroscopic limit.  The lower curve corresponds 
to degenerate quark masses, while the upper curve 
corresponds to an infinite mass splitting.  The 
region between the two curves contains all finite mass splittings. 
The global minimum of the 
energy density for all cases occurs at $\theta=0$.\label{fig:enden}}
\end{figure}

The most fundamental quantity that can be derived from the partition 
function is the energy density which is defined by
\be 
{\cal E}(\theta,\mu_1,\mu_2)=-\frac{1}{V} \, 
	\log\, {\cal Z}^{(N_f=2)}(\theta,\mu_1,\mu_2) \,\, .
\ee
An expansion for large $\mu_{12}(\theta)$ gives 
\be 
{\cal E}(\theta,\mu_1,\mu_2)=-\frac{\mu_{12}(\theta)}{V}
	+ {\cal O}(\log\, \mu_{12}(\theta)) \,\, .
\ee
Note that the volume dependence drops out in the leading order term.
In the macroscopic limit, $\mu_{12}(\theta) \gg 1$,
\be
\lim_{V\to\infty}{\cal E}(\theta,m_1,m_2,V)=
\left\{\begin{array}{l@{\;,\quad}l}
-\Sigma m_1-\Sigma m_2\cos \theta+{\cal O}(m_2/m_1) & m_1\gg m_2 \\
-2m\Sigma|\cos(\theta/2)| & m_1=m_2=m
\end{array}\right.\;
\ee
There is a constant shift, ${\cal E}_{0} = -\Sigma
m_1$, in the energy density which we subtract in the following.
The energy density in both limits is plotted in Fig.\ \ref{fig:enden}. 
It is minimized at $\theta=0$ and maximized at $\theta=\pi$.  For 
degenerate quark masses, there is cusp at $\theta=\pi$.

\subsection{Chiral Condensate}

The chiral condensate is calculated from eq.\ (\ref{eq206}) and 
for the first quark flavor is
\be 
\Sigma^{(N_f=2)}_{(1)}(\theta,\mu_1,\mu_2) = \frac{\Sigma}{2}
\frac{(\mu_1+\mu_2 \cos \theta)}{\mu_{12}(\theta)}
\frac{I_2(\mu_{12}(\theta))}{I_1(\mu_{12}(\theta))}\;.
\ee
The condensate for the second quark flavor is given by the 
same expression with $\mu_1$ and $\mu_2$ interchanged.
The total chiral condensate given by the sum is 
\be
\Sigma^{(N_f=2)}(\theta,\mu_1,\mu_2) = \frac{\Sigma}{2}
\frac{(\mu_1+\mu_2)(1+\cos \theta)}{\mu_{12}(\theta)}
\frac{I_2(\mu_{12}(\theta))}{I_1(\mu_{12}(\theta))}\;.
\ee
For $\theta=0$, the individual contributions 
are equal. For $\theta=\pi$, however, they are equal 
in magnitude but opposite in sign: 
\be
\Sigma^{(N_f=2)}_{(1)}(\pi,\mu_1,\mu_2) = {\rm sign}(\mu_1-\mu_2)
\frac{\Sigma}{2}\frac{I_2(\mu_{12}(\pi))}{I_1(\mu_{12}(\pi))}
         = - \Sigma^{(N_f=2)}_{(2)}(\pi,\mu_1,\mu_2)\;, 
\ee
with their sum, the total chiral condensate, vanishing for any value 
of the quark masses.

Taking of the limit of degenerate quark masses, 
the chiral condensate is 
\bea \label{eq:degSigtwo}
\Sigma^{(N_f=2)}_{(1,2)}(\theta,\mu) &=& 
\Sigma\, \frac{\, \sqrt{2+2\cos \theta}}{4} \, 
\frac{I_2\left(\mu \sqrt{2+2\cos \theta}\right)}{I_1\left(\mu\sqrt{2+2\cos \theta}\right)}\\ \nonumber 
         &=& \frac{\Sigma^{(N_f=2)}(\theta,\mu)}{2}\;.
\eea

\begin{figure}
\centerline{\epsfig{file=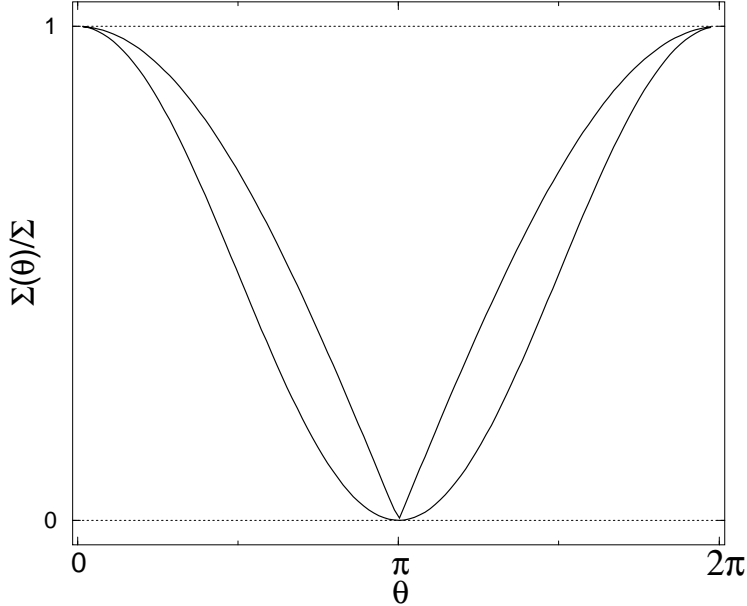,width=10cm}}
\caption{The chiral condensate in the macroscopic limit
as a function of $\theta$ normalized to the infinite 
volume value, $\Sigma$.  The lower curve corresponds 
to asymptotically large quark mass splitting and the upper 
curve is limit of degenerate quark masses.  The area between 
the two curves contains all finite mass splittings between the 
quark masses.  
There is a cusp at $\theta=\pi$ for equal 
quark masses.\label{fig:chirvol}}
\end{figure}

\begin{figure}
\centerline{\epsfig{file=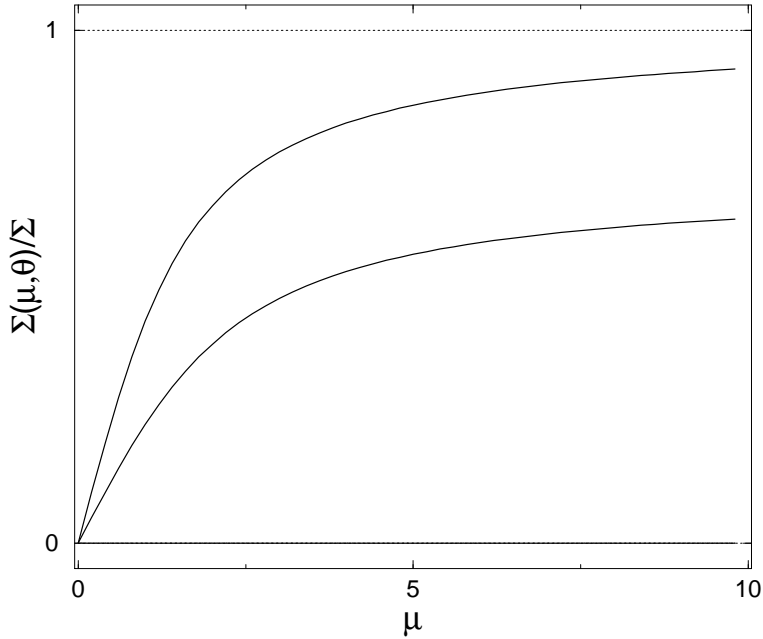,width=10cm}}
\caption{The chiral condensate for $N_f=2$ as a function 
of $\mu$.  The upper, middle and lower curves 
correspond to $\theta=$ 0, $\pi/2$ and $\pi$, 
respectively.  As $\theta$ increases from zero to $\pi$,
the condensate decreases and vanishes identically 
at $\theta=\pi$.\label{fig:chirmuvol}}
\end{figure}

Three observations may be made based upon eq.\ (\ref{eq:degSigtwo})
about the physics of QCD in the Leutwyler--Smilga scaling regime at
nonzero values of $\theta$:

\begin{itemize}

\item For fixed $\mu$, the chiral condensate decreases monotonically 
in the interval $\theta\in [0,\pi)$ and increases monotonically for
$\theta \in (\pi,2\pi)$.

\item As $\mu=m \, \Sigma \, V \rightarrow \infty$, the chiral condensate 
develops a cusp at $\theta=\pi$.

\item For any value of $\mu$, the chiral condensate vanishes identically 
at $\theta=\pi$.

\end{itemize}
This behavior is shown in Figs.\ \ref{fig:chirvol} and \ref{fig:chirmuvol}.  The last observation follows of course from the constancy 
of the partition function for $N_f=$2 at $\theta=\pi$ and is 
not expected to be true outside the Leutwyler--Smilga regime.

For small values of the scaling variable, the chiral condensate has
the expansion
\be 
\frac{\Sigma^{(N_f=2)}(\theta,\mu)}{\Sigma}=\frac{\cos^2(\theta/2)}{2} \,\mu
- \frac{\cos^4(\theta/2)}{12} \,\mu^3+ \frac{\cos^6(\theta/2)}{48} \,\mu^5+{\cal O}(\mu^7) \,\, .
\ee
As expected, the chiral condensate is linear to lowest order in the quark mass and vanishes term-by-term at $\theta=(2n+1)\pi$.  
The expression for the chiral condensate simplifies
in the limit of degenerate quark masses and in the 
limit of large quark mass splitting:
\be
\Sigma^{(N_f=2)}(\theta,m_1,m_2,V)=\left\{\begin{array}{l@{\;,\quad}l}
\Sigma\cos^2(\theta/2)\displaystyle
\frac{I_2(\Sigma V m_1)}{I_1(\Sigma V m_1)} + {\cal O}(m_2/m_1)& m_1\gg m_2 \\ 
\multicolumn{2}{c}{ }\\
\displaystyle\frac{\Sigma}{2} \sqrt{2+2\cos \theta}\, 
\frac{I_2(m \Sigma V \,  \sqrt{2+2\cos \theta})}{I_1(m \Sigma V  \sqrt{2+2\cos \theta})} 
& m_1=m_2=m
\end{array}\right.
\ee

The volume dependence of the chiral condensate in these two limits is
easily calculated.  In the limit of small volume, the chiral
condensate vanishes in both limits of degenerate quark masses and
large mass splittings.  In the macroscopic limit, $V\gg1/\Sigma m$,
the chiral condensate is
\be
\lim_{V\to\infty}\Sigma^{(N_f=2)}(\theta,m_1,m_2,V)=
\left\{\begin{array}{l@{\;,\quad}l}
\Sigma\cos^2(\theta/2) + {\cal O}(m_2/m_1)& m_1\gg m_2 \\
\Sigma|\cos(\theta/2)| & m_1=m_2=m
\end{array}\right.\;
\ee
and is independent of the quark masses in both cases.  Any nonzero
quark mass splitting, $m_1\neq m_2$, interpolates between these two
cases as shown in Fig.\ \ref{fig:chirvol}.  The value of the chiral
condensate is remarkably insensitive to the magnitude of the quark
mass splitting. 

In Fig.\ \ref{fig:chirmuvol}, the chiral condensate as a function of
the scaling variable is plotted for $\theta=$ 0, $\pi/2$, and $\pi$.
As $\theta$ increases from zero to $\pi$, the condensate decreases
monotonically and vanishes identically at $\theta=\pi$.  The monotonic
decrease in the chiral condensate can be understood from the
underlying QCD Lagrangian \cite{poulpriv}.  If one performs a chiral
rotation by an angle $\alpha$, then $\theta \rightarrow \theta-\alpha$
and the chiral condensate, $\langle {\bar \psi} \psi \rangle$ rotates
into $\langle {\bar \psi} \gamma_5 \psi\rangle$.  For $N_f=2$, the
(scalar) chiral condensate is completely rotated into the pseudoscalar
chiral condensate at $\theta=\pi$.  This can be demonstrated
analytically by introducing a complex mass matrix as a source for both
the scalar and pseudoscalar chiral condensates.

\subsection{Chiral Susceptibility}

The chiral susceptibility has been previously studied in Ref.\
\cite{Smilga:1995tb} but the analysis was restricted to 
fixed values of $\nu$.  As shown in Ref.\ \cite{Damgaard1}, 
however, the contributions from all topological sectors 
can be important in the mesoscopic regime. 
Defining
\begin{mathletters}
\bea
{\cal S}(\mu_{12}(\theta)) & = &
\left[\frac{I_3(\mu_{12}(\theta))}{I_1(\mu_{12}(\theta))}-
\frac{I_2(\mu_{12}(\theta))^2}{I_1(\mu_{12}(\theta))^2}  \right] \\
{\cal T}(\mu_{12}(\theta)) & = &
\frac{1}{\mu_{12}(\theta)}
\frac{I_2(\mu_{12}(\theta))}{I_1(\mu_{12}(\theta))}\;,
\eea
\end{mathletters}
the diagonal terms in the chiral susceptibility are 
\begin{mathletters}
\bea 
\chi_{11} &=& \Sigma^2V\left[
        \frac{(\mu_1+\mu_2 \cos \theta)^2}{\mu_{12}(\theta)^2}
        {\cal S}(\mu_{12}(\theta))
        +{\cal T}(\mu_{12}(\theta))\right]\\
\chi_{22} &=& \Sigma^2 V\left[
        \frac{(\mu_2+\mu_1 \cos \theta)^2}{\mu_{12}(\theta)^2}
        {\cal S}(\mu_{12}(\theta))
        +{\cal T}(\mu_{12}(\theta))\right]\,\, .
\eea
\end{mathletters}
The off--diagonal terms are 
\be
        \chi_{12} = \chi_{21}= 
\Sigma^2 V\left[\frac{(\mu_1+\mu_2\cos \theta)
(\mu_2+\mu_1\cos \theta)}{\mu_{12}(\theta)^2}
{\cal S}(\mu_{12}(\theta))
+\cos \theta{\cal T}(\mu_{12}(\theta))\right]\;.  
\ee
In the limit of degenerate quark masses, $m_1=m_2=m$, chiral
susceptibilities become
\begin{mathletters}
\bea
\chi_{11} = \chi_{22} &=& \frac{\Sigma}{m}\left[\mu\cos^2(\theta/2)
        {\cal S}(\mu(\theta))+\mu{\cal T}(\mu(\theta))\right]\\
\chi_{12} = \chi_{21} & = &
        \frac{\Sigma}{m}\left[\mu\cos^2(\theta/2)
        {\cal S}(\mu(\theta))
        +\mu\cos \theta{\cal T}(\mu(\theta))\right]\;.
\eea
\end{mathletters}
At $\theta=\pi$, the diagonal and off-diagonal elements are
maximally different,
\begin{mathletters}
\bea
\chi_{11}(\theta=\pi)=\chi_{22}(\theta=\pi) 
&=& \frac{\Sigma\mu}{4m}=\frac{\Sigma^2 V}{4} \\ 
\chi_{12}(\theta=\pi)=\chi_{21}(\theta=\pi) 
&=&-\frac{\Sigma\mu}{4m}= -\frac{\Sigma^2 V}{4}\;.
\eea
\end{mathletters}
An expansion in small $\mu$ yields
\begin{mathletters}
\bea
\chi_{11} = \chi_{22} &=& \frac{\Sigma}{m} \left[
\frac{\mu}{4}-\frac{1}{24}\cos^2(\theta/2)(2+\cos \theta)\mu^3
+{\cal O}(\mu^5)\right] \\
\chi_{12} = \chi_{21} &=& \frac{\Sigma}{m}\left[
\frac{\mu\cos \theta}{4}-\frac{1}{24}\cos^2(\theta/2)(1+2\cos \theta)\mu^3
+{\cal O}(\mu^5)\right]\;.
\eea
\end{mathletters}
On the other hand, in the limit $\mu\gg1$ for $\theta\neq\pi$,
\begin{mathletters}
\bea \label{eq:chivsa}
\chi_{11}=\chi_{22} &=& \frac{\Sigma}{m}\left[
\frac{\sin^2(\theta/2)}{2\left|\cos(\theta/2)\right|} 
+\frac{3\cos \theta}{8 \mu \cos(\theta/2)^2} 
+{\cal O}(1/\mu^2)\right]\\
\chi_{12}=\chi_{21} &=& - \frac{\Sigma}{m}\left[
\frac{\sin^2(\theta/2)}{2\left|\cos(\theta/2)\right|} \label{eq:chivsb}
+\frac{3}{8 \mu \cos(\theta/2)^2}+{\cal O}(1/\mu^2)\right]\; ,
\eea
\end{mathletters}
therefore, to leading order in $1/\mu$,
$\chi_{11}=\chi_{22}=-\chi_{12}=-\chi_{21}$.  The chiral
susceptibility was shown to be proportional to $3/8m^2 V$ at
$\theta=0$ in Ref.\ \cite{Smilga:1995tb}.  The importance of 
summing over all topological charges is clear in this expression, 
since the leading order terms in eqs.\ (\ref{eq:chivsa}) and (\ref{eq:chivsb})
vanish at $\theta=0$.

\subsection{Topological Density: First--Order Phase Transition at ${\fett \theta=\pi}$}

The topological density, defined by eq.\ (\ref{eq207}), is
\be \label{eq:sigtwo}
\sigma(\theta,\mu_1,\mu_2) 
= \frac{1}{V} \, \frac{\mu_1 \mu_2 \sin \theta}{\mu_{12}(\theta)} 
\frac{I_{2}(\mu_{12}(\theta))}{I_{1}(\mu_{12}(\theta))} \;.
\ee
At $\theta=0$, $\sigma(\theta,\mu_1,\mu_2) $ has the strict
interpretation as the mean topological charge.  Since in a large enough
ensemble of gauge fields the average number of instantons should be
equal to average number of anti-instantons, the topological density
should be zero at $\theta=0$, and, indeed, the right hand side of eq.\
(\ref{eq:sigtwo}) vanishes for $\theta$ equal to any integer multiple
of $\pi$.

In the limit of degenerate quark masses, the topological density becomes
\be
\sigma(\theta,\mu) = \frac{1}{V} \, \frac{\mu\sin \theta}{\sqrt{2+2\cos \theta}}
        \, \frac{I_{2}\left( \mu \sqrt{2+2\cos \theta}\right)}{I_{1}\left(\mu \sqrt{2+2\cos \theta}\right)} \;.
\ee
The most interesting property of this relation is that 
in the limit of very large scaling variable, $\mu \gg 1$,
$\sigma(\theta,\mu)$ 
develops a discontinuity at $\theta=\pi$.  This is the
first--order phase transition proposed by Dashen \cite{Dashen}.

\be
\sigma(\theta,m_1,m_2,V)=\left\{\begin{array}{l@{\;,\quad}l}
\Sigma m_2\sin \theta\displaystyle
\frac{I_2(\Sigma V m_1)}{I_1(\Sigma V m_1)}
+{\cal O}(m_2/m_1)  & m_1\gg m_2   \\
\multicolumn{2}{c}{ }\\
\Sigma m \sin(\theta/2)\displaystyle
\frac{I_2(2\Sigma V m\cos(\theta/2))}{I_1(2\Sigma V m\cos(\theta/2))} 
& m_1=m_2=m
\end{array}\right.
\ee
Then, in the infinite volume limit,
\be
\lim_{V\to\infty}\sigma(\theta,m_1,m_2,V)=
\left\{\begin{array}{l@{\;,\quad}l}
\Sigma m_2\sin \theta+{\cal O}(m_2/m_1) & m_1\gg m_2   \\
\Sigma m \sin(\theta/2) {\rm sign}(\cos(\theta/2))& m_1=m_2=m \,\, .
\end{array}\right.
\ee
The limit $m_1\gg m_2$ coincides with the one-flavor case.
For two flavors of degenerate quarks, there is a first--order phase
transition at $\theta=\pi$, while for two flavors of non-degenerate
quarks the phase transition disappears as long as the quark masses 
are kept finite.  As mentioned above, it was shown in 
Ref.\ \cite{Akemann:2001ir} that there exists 
a particular scaling limit in which there may be a phase 
transition for nondegenerate quark masses.

\begin{figure}
\centerline{\epsfig{file=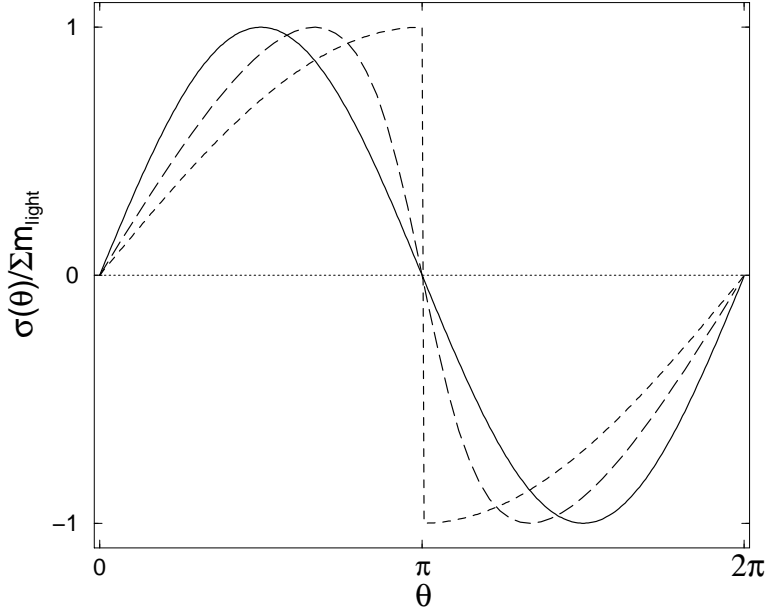,width=10cm}}
\caption{The topological density of the $\theta$ vacua. 
The solid line corresponds to the limit where one quark mass is very
heavy, $m_1\gg m_2$, the long dashed line to mass splitting of
$m_1/m_2=2$, and the dashed line to the case of degenerate masses,
$m_1=m_2=m$. A first-order phase transition occurs at $\theta=\pi$ in the case
of degenerate quark masses.  The transition is washed out by mass
splitting between the quarks. The topological density is measured in units
of $m_2\Sigma$, where $m_2$ denotes the mass of the lighter quark.
\label{figep}}
\end{figure}

\subsection{Topological Susceptibility}

The topological susceptibility reads
\bea
\lefteqn{\chi^{(N_f=2)}(\theta,m_1,m_2,V)} 
\\ \nonumber & = & \frac{1}{V}
\frac{I_2(\mu_{12}(\theta))}{I_1(\mu_{12}(\theta))}
\frac{\mu_1^2\mu_2^2}{\mu_{12}^2(\theta)}
\left(\frac{\mu_{12}(\theta)\cos \theta}{\mu_1\mu_2}
+\left(\frac{4}{\mu_{12}(\theta)}
+\frac{I_2(\mu_{12}(\theta))}{I_1(\mu_{12}(\theta))}
-\frac{I_1(\mu_{12}(\theta))}{I_2(\mu_{12}(\theta))}\right)
\sin^2(\theta)\right)\;.
\eea
For equal quark masses, one has
\bea
\chi^{(N_f=2)}(\theta,m,V) = 
\frac{\Sigma m}{2}\frac{2-\cos \theta}{\left|\cos(\theta/2)\right|}
\frac{I_2(\mu(\theta))}{I_1(\mu(\theta))}
+ \frac{\Sigma^2 m^2 V}{4}\frac{\sin^2(\theta)}{\cos^2(\theta/2)}
\left(\frac{I_2(\mu(\theta))^2}{I_1(\mu(\theta))^2}
-1\right)\;.
\eea
This reduces at $\theta=\pi$ to
\be
\chi^{(N_f=2)}(\theta=\pi,m,V)=-\frac{\Sigma^2m^2}{4}V\;,
\ee
which diverges in the infinite volume limit on account of the 
first-order phase transition. For $\theta\neq\pi$,
however, the limit is finite,
\be \label{eq:wardtak}
\lim_{V\to\infty}\chi^{(N_f=2)}(\theta,m,V)
=\frac{\Sigma m}{2}|\cos(\theta/2)|\;,\quad\theta\neq\pi\;.
\ee
This is completely consistent with the expectation from 
the flavor singlet Ward--Takahashi identity at $\theta=0$ which 
predicts a linear rise in topological susceptibility with mass
\cite{Crewther:1977ce},
\be \label{WardT}
\lim_{V\to\infty}\chi^{(N_f)}(\theta=0,m,V)
=\frac{\Sigma m}{N_f} + {\cal O}(m^2)\,\, .
\ee
Equation (\ref{eq:wardtak}) may be considered to be a generalization 
of eq.\ (\ref{WardT}) to nonzero values of $\theta$.

In the limit of large mass splitting, the topological susceptibility reduces
to 
\be
\chi^{(N_f=2)}(\theta,m_1,m_2,V)=\Sigma \, m_2 \, \cos \theta
\frac{I_2(\Sigma V m_1)}{I_1(\Sigma V m_1)}
+{\cal O}(m_2/m_1)\;,\quad m_1\gg m_2\;,
\ee
which in the macroscopic limit is 
\be
\lim_{V\to\infty}\chi^{(N_f=2)}(\theta,m_1,m_2,V)
=\Sigma \, m_2 \, \cos \theta
+{\cal O}(m_2/m_1)\;,\quad m_1\gg m_2\; .
\ee

\begin{figure}
\centerline{\epsfig{file=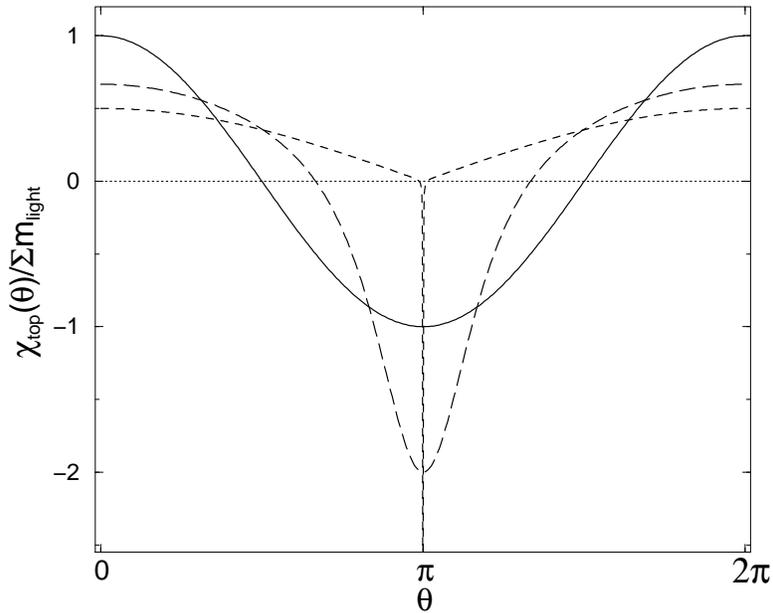,width=10cm}}
\caption{The topological susceptibility as a function $\theta$ in the 
macroscopic limit. The solid, long dashed, and dashed line represent
$m_1\gg m_2$, $m_1/m_2=2$, and $m_1=m_2$, respectively.}
\end{figure}


\section{Vacuum Properties for ${\fett N_f=3}$} \label{sec:threeflavors}

The summation over all topological charges in the partition function for
$N_f \ge 3$ was performed in Sec.\ \ref{sec:sum}.  In this section, we
complement the analysis of Sec.\ \ref{sec:twoflavors} by examining the
$\theta$ dependence of the QCD vacuum for three flavors of quarks.
The partition function for $N_f=3$ involves a single integration which
we were not able to calculate analytically.  Since the partition
function has a group--theoretic origin, we expect that the integrals
that result after summing over all topological charges provide some
functional representation of the Goldstone manifold.  However, since
the integrand is a smooth function of $\theta$ and $\mu$ except
possibly at a single point in the interval of integration and the
integral is taken over a compact interval, the numerical evaluation of
the partition function is straightforward.  Additionally, the
derivatives of the partition function with respect to $\theta$ and
$\mu$ can be commuted past the integration.  Explicit expressions are
only given for the chiral condensate and the topological density since
the expressions for the chiral and topological susceptibilities become
quite complicated and are not particularly enlightening even in the
limit of degenerate quark masses.  We focus on a triplet of quark
masses, $(m_{\rm light},m_{\rm light},m_{\rm heavy})$, between the
limits of total degeneracy, $m_{\rm light}=m_{\rm heavy}$, and very
large mass splitting, $m_{\rm light} \ll m_{\rm heavy}$.

\subsection{Partition Function}

For three degenerate masses, the partition function can be written in
the compact form:
\bea 
{\cal Z}^{(N_f=3)}(\theta,\mu) &=& \frac{2}{\pi}\int\limits_0^{2\pi} d\phi \nonumber
        \frac{e^{\mu\cos \phi}}{\mu(\theta,\phi)}
        \left[ \cos\left(\frac{3\phi-\theta}{2}\right)^2
        \, I_1(\mu(\theta,\phi)) - 
        \cos\left(\frac{3\phi-\theta}{2}\right)
        \, I_2(\mu(\theta,\phi))\right] \\ \nonumber
\vspace{2cm} \\  
&=& \frac{2}{\pi} \int\limits_0^{2\pi} d\phi 
        \frac{e^{\mu \cos \phi}}{\mu(\theta,\phi)}
        \left| 
        \begin{array}{cc}
        I_1(\mu(\theta,\phi)) & I_2(\mu(\theta,\phi)) \\
        \cos\left(\frac{3\phi-\theta}{2}\right) & 
        \cos\left(\frac{3\phi-\theta}{2}\right)^2 
        \end{array} 
        \right| \,\, , 
\eea
where
$\mu(\theta,\phi) = 2 \mu \left|\cos\left(\frac{\phi-\theta}{2}\right)\right|$.  Unlike the $N_f=2$ partition function, we find that ${\cal
Z}^{(N_f=3)}(\theta,\mu)$ does not become independent of $\mu$ at
$\theta=\pi$ which is consistent with the discussion in Sec.\ \ref{spectral}.  As discussed in Refs.\ \cite{Smilga:1999dh,Tytgat:1999yx}, 
the leading order term 
in the chiral expansion does not vanish at $\theta=\pi$ for 
$N_f\ge$3.

\begin{figure}
\centerline{\epsfig{file=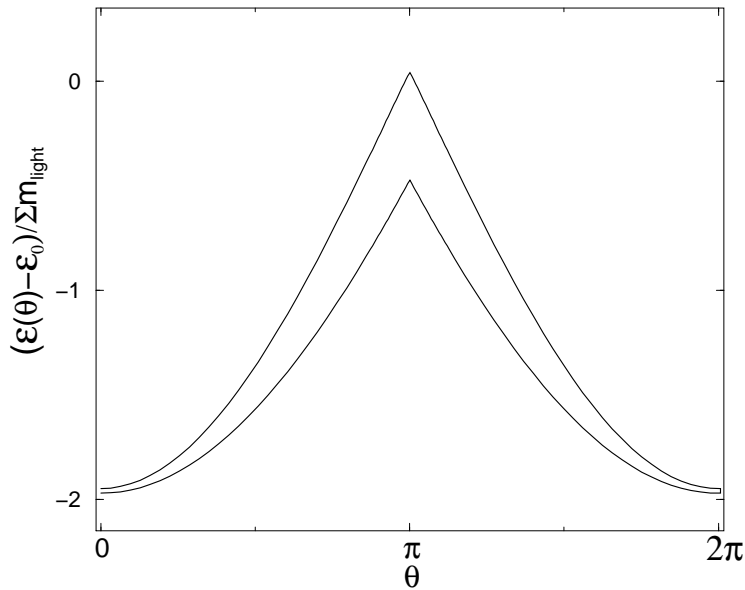,width=10cm}}
\caption{The energy density in the macroscopic limit as a function of
$\theta$.  The upper (lower) curve corresponds to infinite (zero) mass
splitting between the two degenerate quark masses and the third
quark mass.  All finite mass splittings are contained in
the region between the two curves.\label{fig:3energy}}
\end{figure}

The dependence of the energy density as a function of $\theta$,
however, is similar to the two flavor case.  The energy density in the
macroscopic limit is shown in Fig.\ \ref{fig:3energy} shifted by its
zero point, ${\cal E}_0 = -\Sigma m_{\rm heavy}$.  The
global minimum and maximum of the energy density in the macroscopic
limit are at $\theta=0$ and $\pi$, respectively.  As long as two quark
masses are degenerate, there seems to be cusp at $\theta=\pi$.  In
general, the positions for the global minimum and maximum are
$\theta=0$ and $\pi$, respectively, for any value of $\mu$.  The zero point 
energy is dominated by the heavy quark mass.  After subtracting 
this physically irrelevant zero point contribution, the energy 
density is relatively insensitive to taking the heavy quark mass 
to infinity.

\subsection{Chiral Condensate}

\begin{figure}
\centerline{\epsfig{file=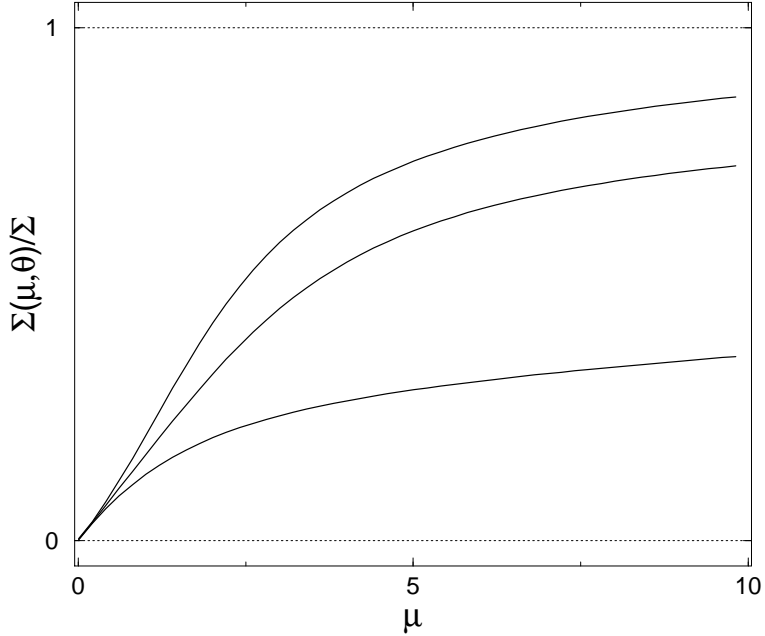,width=10cm}}
\caption{The mass dependence of the chiral condensate for various
values of $\theta$ for three degenerate quark flavors.  The upper,
middle and lower curves correspond to $\theta=0$, $\pi/2$, and $\pi$,
respectively.  As $\theta$ is increased from zero to $\pi$, the chiral
condensate decreases in magnitude for a given $\mu$.\label{fig:3cond}}
\end{figure}

For three degenerate quark masses, the chiral condensate is 
\begin{mathletters}
\bea 
\Sigma^{(N_f=3)}(\theta,\mu) &=& \frac{2}{\pi}\,\frac{\Sigma}{{\cal Z}^{(N_f=3)}(\theta,\mu)} 
        \int\limits_{0}^{2\pi} \, d\phi
        \frac{e^{\mu\cos \phi}\,}
        {\mu(\theta,\phi)}
        \Bigg[ a_1\, 
	I_1(\mu(\theta,\phi)) 
	+  a_2\, 
	I_2(\mu(\theta,\phi))\Bigg] \,\, ,
\eea
where 
\bea 
a_1 &=& \cos\left(\frac{3\phi-\theta}{2}\right)
	\left[\cos\phi \, \cos\left(\frac{3\phi-\theta}{2}\right)-
	\left|\cos\left(\frac{\phi-\theta}{2}\right)\right|\right] \\ 
a_2 &=& \cos\left(\frac{3\phi-\theta}{2}\right)
	\left[\frac{3}{\mu}-\cos \phi + 
	2 \cos\left(\frac{3\phi-\theta}{2}\right)
	\left|\cos\left(\frac{\phi-\theta}{2}\right)\right| \right] \,\, .
\eea
\end{mathletters}
In many respects, the behavior of the equal mass chiral condensate for
two and three quark flavors is again similar.  The chiral condensate as a
function of the scaling variable is plotted in Fig.\ \ref{fig:3cond}.
The chiral condensate decreases monotonically as $\theta$ increases
from zero to $\pi$ like the $N_f=2$ case.  However, it is nonzero at
$\theta=\pi$ unlike the $N_f=2$ chiral condensate.  For small values 
of $\mu$, the chiral condensate is linear in $\mu$ and independent of 
$\theta$.
 
\begin{figure}
\centerline{\epsfig{file=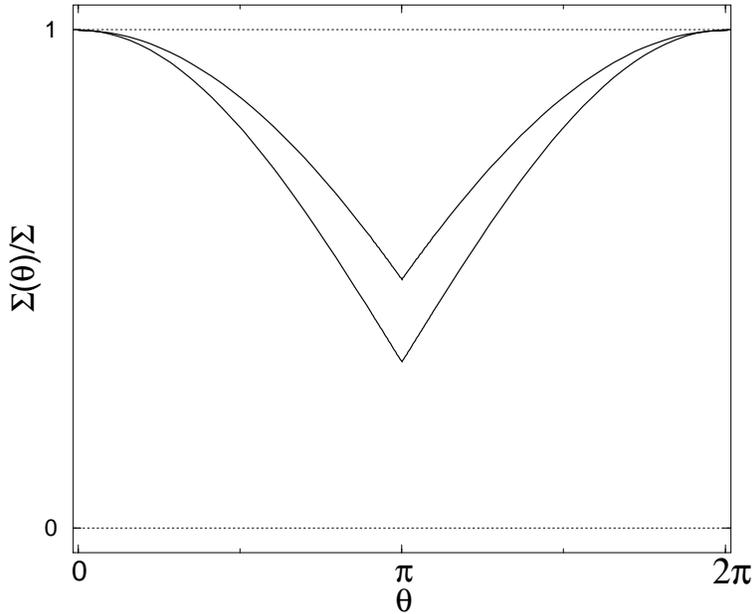,width=10cm}}
\caption{The three-flavor chiral condensate in the macroscopic 
limit as a function of $\theta$ for which two of the quark masses 
are degenerate.  The upper (lower) curve corresponds 
to infinite (zero) mass splitting between the two 
degenerate quark masses and the third quark mass.
All finite mass splittings are in the region between the two curves.
\label{fig:condnf3}}
\end{figure}

Figure \ref{fig:condnf3} shows the chiral condensate as a function of
$\theta$ in the macroscopic limit, $\mu \rightarrow \infty$.  In this
limit, $\Sigma^{(N_f=3)}(\theta,\mu)$ has a cusp at $\theta=\pi$, but
otherwise is a smooth function.  In the macroscopic limit with two
degenerate quark masses, the value of the chiral condensate is very
insensitive to the mass of the third nondegenerate quark.  For 
any nonzero value of $\theta$, the chiral condensate is always 
less than the $\theta=0$ chiral condensate.  We find numerically 
that 
\be
\frac{\Sigma(\theta=\pi,m_{\rm light},m_{\rm heavy})}{\Sigma} = 
\left\{\begin{array}{l@{\;,\quad}l}
\displaystyle\frac{1}{2}& m_{\rm light}=m_{\rm heavy}   \\
\multicolumn{2}{c}{ }\\
\displaystyle\frac{1}{3}& m_{\rm light}\ll m_{\rm heavy}
\end{array}\right.
\ee
to a very high precision.

\subsection{Topological Density: First-Order Phase Transition at ${\fett \theta=\pi}$}

\begin{figure}
\centerline{\epsfig{file=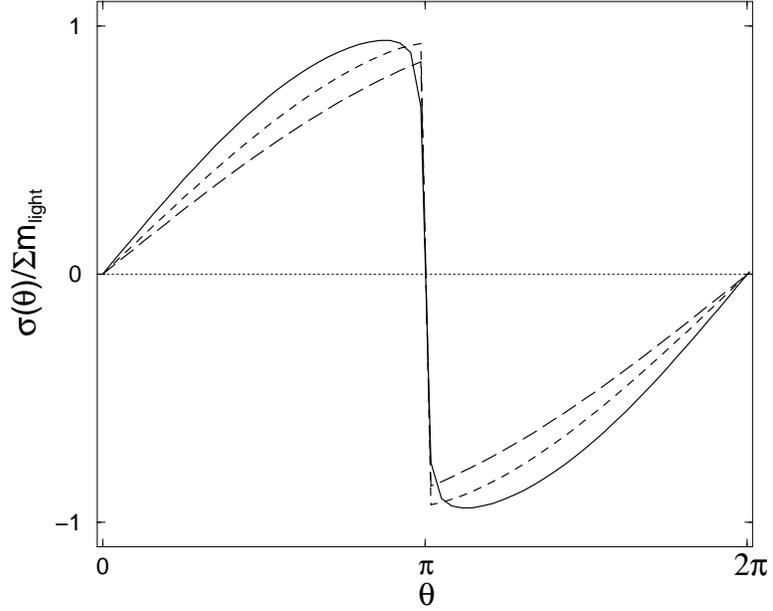,width=10cm}}
\caption{The $N_f=3$ topological density 
as a function of $\theta$ in the macroscopic limit
for which at least two of the quark masses are degenerate. 
The solid, dashed and long--dashed curves correspond to 
the mass splittings $m_{\rm heavy}\gg m_{\rm light}$, 
$m_{\rm heavy}= 2 m_{\rm light}$, and $m_{\rm heavy}=m_{\rm light}$, 
respectively.
For completely degenerate masses, there is a first--order 
phase transition at $\theta=\pi$.  Even for large mass 
splittings, however, there is still a very rapid crossover.
\label{fig:3topden}}
\end{figure}

For degenerate quark masses, the topological density 
is given by
\begin{mathletters}
\bea 
\sigma(\theta,\mu,V) &=& \frac{2}{\pi} \, 
	\frac{1}{{\cal Z}^{(N_f=3)}}\, \frac{1}{V}
        \int\limits_{0}^{2\pi} \, d\phi \,  
	\frac{e^{\mu\cos \phi} }
	{\mu(\theta,\phi)}\,
        \Bigg[ b_1\, 
	I_1\left(\mu(\theta,\phi)\right)
	+ b_2 \, I_2\left(\mu(\theta,\phi)
	\right)\Bigg]
\eea
where the coefficients are given by 
\bea
b_1 &\equiv& \cos\left(\frac{3\phi-\theta}{2}\right)
	\left[-\mu\, \sin\left(\frac{\phi-\theta}{2}\right)+ 
	\sin\left(\frac{3\phi-\theta}{2}\right)\right]\\ 
b_2 &\equiv& \cos\left(\frac{3\phi-\theta}{2}\right)
	\left[\mu\, \cos\left(\frac{3\phi-\theta}{2}\right)
	\sin\left(\frac{\phi-\theta}{2}\right)+\frac{3}{2} 
	\tan\left(\frac{\phi-\theta}{2}\right)-
	\frac{1}{2}\tan\left(\frac{3\phi-\theta}{2}\right)\right] \,\, .
\eea
\end{mathletters}
The topological density as a function of $\theta$ is shown in Fig.\
\ref{fig:3topden} in the macroscopic limit.  For three degenerate
quark masses, there is a discontinuity at $\theta=\pi$ which is simply
Dashen's phenomena.  This discontinuity is washed out for any nonzero
mass splitting, however, for any mass splitting between $m_{\rm heavy}$
and $m_{\rm light}$, the transition is always extremely rapid.

\subsection{Topological Susceptibility} 

\begin{figure}
\centerline{\epsfig{file=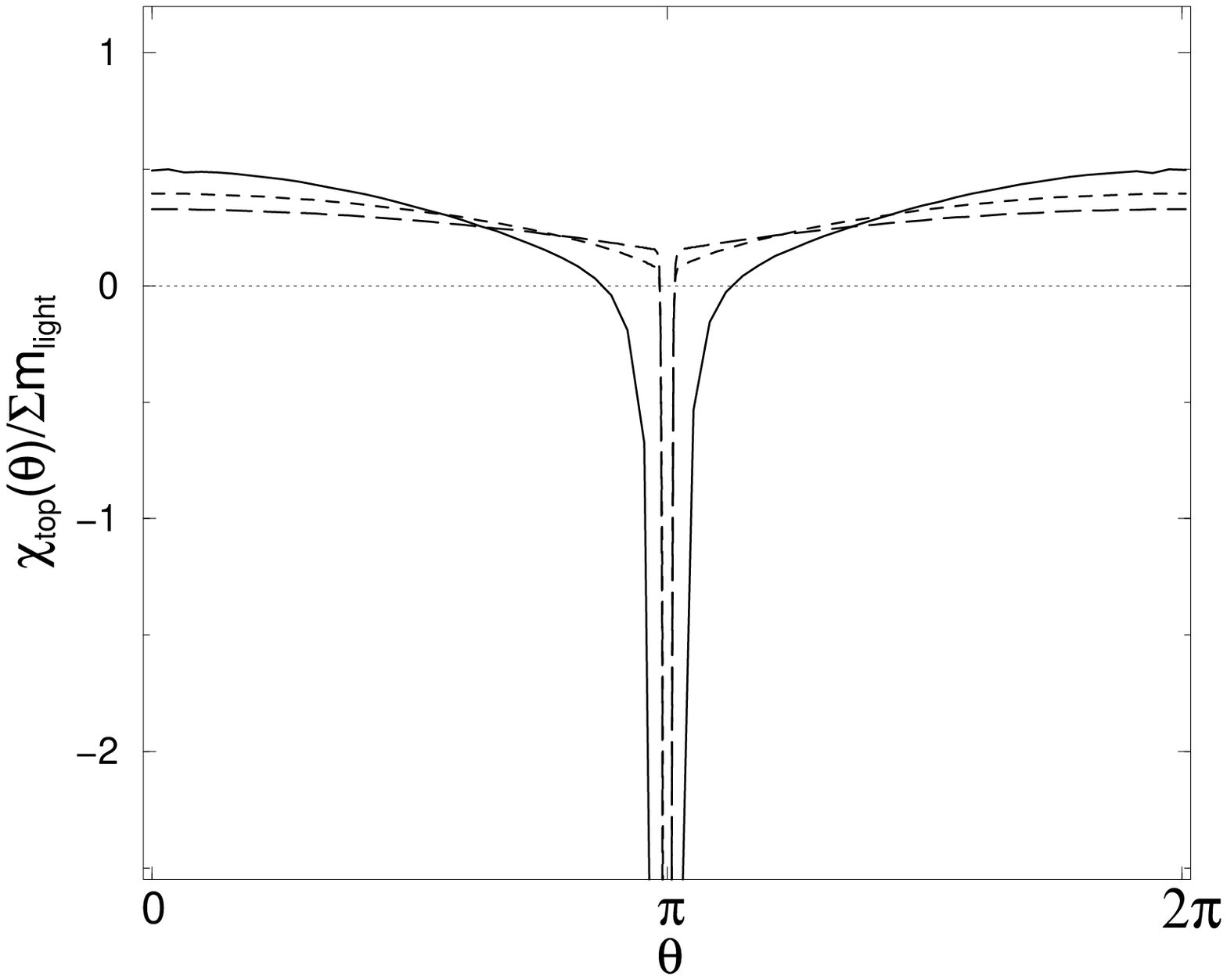,width=10cm}}
\caption{The $N_f=3$ topological susceptibility as a function of
$\theta$ in the macroscopic limit for which at least two of the quark
masses are degenerate.  The solid, dashed and long--dashed curves
correspond to the mass splittings $m_{\rm heavy}\gg m_{\rm light}$,
$m_{\rm heavy}= 2 m_{\rm light}$, and $m_{\rm heavy}=m_{\rm light}$, respectively.
For completely degenerate masses, the topological susceptibility
diverges at $\theta=\pi$ on account the first-order phase transition.
\label{fig:topsusc_nf3}}
\end{figure}

The topological susceptibility in the macroscopic limit for $N_f=3$ is
shown in Fig.\ \ref{fig:topsusc_nf3}.  For three degenerate masses,
the topological susceptibility diverges at $\theta=\pi$ on account of
the first-order phase transition.  For three quark flavors with only
two degenerate masses, there is still an extreme drop in the
topological susceptibility at $\theta=\pi$ even for a very large mass
splitting.  For two degenerate quark flavors, the topological
susceptibility is positive except in the vicinity of $\theta =\pi$.
We find numerically that
\be
\lim_{V\rightarrow\infty} \chi^{(N_f=3)}
(\theta=0,m_{\rm light},m_{\rm heavy},V) = 
\left\{\begin{array}{l@{\;,\quad}l}
\displaystyle\frac{\Sigma \, m_{\rm light}}{3}& m_{\rm light}=m_{\rm heavy}   \\
\multicolumn{2}{c}{ }\\
\displaystyle\frac{\Sigma \, m_{\rm light}}{2}& m_{\rm light}\ll m_{\rm heavy}
\end{array}\right.
\ee
which is consistent with the Ward--Takahashi identity (\ref{WardT}).

\section{Conclusion} \label{sec:theend}

In this paper, we investigated the properties of the QCD partition
function in the Leutwyler--Smilga finite volume scaling regime.  The
full partition function including the contributions from all
topological sectors for $N_f=2$ has been previously calculated.  We
extended these results by performing the summation over all
topological charges for arbitrary $N_f$.  For $N_f\ge3$, the partition
function can be expressed as a $(N_f-2)$--fold angular integration 
over a
finite sum of modified Bessel functions.  We considered both the
cases of degenerate and nondegenerate quark masses.  The partition 
function remains $2\pi$-periodic in $\theta$ after summing 
over all topological charges.

We systematically investigated the $\theta$ dependence of the QCD
vacuum in the Leutwyler--Smilga regime.  In this limit, the partition
function only depends on $\theta$ and the scaling variables, $\mu_i=m_i
\Sigma V$.  In the limit of degenerate quark masses, the $N_f=2$
partition function is independent of the scaling variable at
$\theta=\pi$.  As a consequence, the chiral condensate vanishes
identically at $\theta=\pi$ for all values of $\mu$.  
For fixed $\mu$,
the chiral condensate decreases monotonically as $\theta$ is increased
from $0$ to $\pi$.

In the macroscopic limit, i.e. for $\mu\rightarrow\infty$, the
behavior of the two-flavor partition function is particularly
striking.  For degenerate quark masses, the first derivative of the
energy density with respect to $\theta$ has a discontinuity at
$\theta=\pi$ corresponding to the spontaneous breaking of the discrete
CP symmetry.  This phenomena was predicted by Dashen and was
subsequently demonstrated by Di Vecchia and Veneziano and Witten using
large-$N_c$ chiral perturbation theory.  The chiral condensate also
develops a cusp at $\theta=\pi$ for degenerate quark masses in the
macroscopic limit.

For $N_f=3$, we find that all the examined 
quantities are very insensitive to the mass splitting between 
two degenerate flavors and the third heavy flavor in the 
macroscopic limit.  When compared to those of $N_f=2$,
the vacuum properties for three flavors are quite similar.  While the
partition function for $N_f=2$ can be expressed in closed form, the
partition function for $N_f=3$ requires the straightforward 
numerical evaluation of a 
single integral.
Unlike the $N_f=2$ case, the $N_f=3$ partition function is not
independent of the scaling variables at $\theta=\pi$ and subsequently
the chiral condensate is nonzero at $\theta=\pi$ even in the
macroscopic limit.  The chiral condensate does, however, exhibit the
same monotonicity in $\theta$ as the $N_f=2$ chiral condensate and in
the macroscopic limit there is a cusp at $\theta=\pi$.  Dashen's
phenomena, a first-order phase transition at $\theta=\pi$, is also
realized in this limit.

Many of our results are corroborated by general $\theta=0$ field
theoretic identities, for example, the Ward--Takahashi identity.  Our
analysis naturally extends these results to nonzero values of
$\theta$.  We have also demonstrated that QCD in the mesoscopic regime
exhibits Dashen's phenomena.  We examined the physics of QCD at
$\theta \neq 0$ in a nonperturbative framework.  While the existence
of Dashen's phenomena has been known for thirty years and studied
using chiral perturbation theory and numerical simulations, we have
demonstrated the spontaneous breaking of the CP symmetry at
$\theta=\pi$ in a way that is both nonperturbative and analytic.  We
have also provided a further step towards the full evaluation of the
path integral for chiral perturbation theory.  We hope that this work
helps elucidate a parameter space of QCD which has been largely
unexplored.

\begin{center} 
{\bf Acknowledgments} 
\end{center}

We especially thank P.H.\ Damgaard for very useful comments and
suggestions.  We acknowledge D.\ Diakonov for posing a question that
led to the present work and thank G.\ Akemann, 
S.\ Chandrasekharan, C.\ Dawson, T.\
Fugleberg, A.\ Jackson, D.\ Kharzeev, M.\ Oswald, O.\ Philipsen, R.\
Pisarski and K.\ Splittorff for useful conversations and suggestions.

\appendix

\section{Explicit Calculation of the Degenerate Mass Partition Function}
\label{app1}

Starting with the representation 
\begin{equation}
{\cal Z}^{(N_f)}_\nu(\mu)=
\int\limits_0^{2\pi}\prod_{j=1}^{N_f} \left(\frac{d\phi_j}{2\pi}
\exp\left[\mu\cos(\phi_j)-i\nu\phi_j-ij\phi_j \right]\right)
\varepsilon_{i_1\ldots i_{N_f}} e^{i\phi_{i_1}}\ldots e^{iN_f\phi_{i_{N_f}}}
\;,
\end{equation}
the summation ${\cal Z}^{(N_f)}(\theta,\mu)=\sum_\nu
e^{i\nu\theta}{\cal Z}^{(N_f)}_\nu(\mu)$ can be performed by the use of
\begin{equation}
\int_0^{2\pi}\frac{d\phi_1}{2\pi} f(\phi_1,\ldots,\phi_{N_f},\theta)
\sum_{\nu=-\infty}^\infty e^{i\nu(\theta-\phi_1\ldots-\phi_{N_f})}=
f(\phi_2+\ldots+\phi_{N_f}-\theta,\phi_2,\ldots,\phi_{N_f},\theta)\;,
\end{equation}
After a change of integration variables
\bea
e^{i\phi_2}(1+e^{i(\phi_3+\ldots+\phi_{N_f}-\theta)})&=&\psi e^{i\varphi}\\
e^{-i\phi_2}(1+e^{-i(\phi_3+\ldots+\phi_{N_f}-\theta)})&=&\psi e^{-i\varphi}
\label{eq427}
\eea
with
\be
\psi=\sqrt{2+2\cos\left(\phi_3+\ldots+\phi_{N_f}-\theta\right)}
\label{eq428}
\ee
one arrives at
\bea
\label{eq429}
\lefteqn{{\cal Z}^{(N_f)}(\theta,\mu)}
\\ \nonumber & = &
e^{-i\theta/2}\int\limits_0^{2\pi}\frac{d\varphi}{2\pi}
e^{\mu\psi\cos\varphi-i\varphi} \int\limits_0^{2\pi} 
\prod_{j=3}^{N_f}\left(\frac{d\phi_j}{2\pi}
e^{\mu\cos(\phi_j)-i(j-3/2)\phi_j}\right)
\Gamma(\varphi,\phi_3,\ldots,\phi_{N_f},\theta)
\eea
where
\be
\Gamma(\varphi,\phi_3,\ldots,\phi_{N_f},\theta)=
\left|\begin{array}{cccc}
1 & e^{-\frac{i}{2}(\phi_3\ldots+\phi_{N_f}-\theta)-i\varphi} &
\cdots & e^{-\frac{i(N_f-1)}{2}(\phi_3\ldots+\phi_{N_f}-\theta)-i(N_f-1)\varphi} \\
1 & e^{-\frac{i}{2}(\phi_3\ldots+\phi_{N_f}-\theta)+i\varphi} &
\cdots & e^{-\frac{i(N_f-1)}{2}(\phi_3\ldots+\phi_{N_f}-\theta)+i(N_f-1)\varphi} \\
1 & e^{i\phi_3} & \cdots & e^{i(N_f-1)\phi_3} \\
\vdots & \vdots & \ddots & \vdots \\
1 & e^{i\phi_{N_f}} & \cdots & e^{i(N_f-1)\phi_{N_f}}
\end{array}\right|\;.
\label{eq430}
\ee
Expanding the determinant around the first two rows leads to
\be 
\Gamma(\varphi,\phi_3,\ldots,\phi_{N_f},\theta)=
\sum_{r=1}^{N_f-1}(-1)^r(e^{ir\varphi}-e^{-ir\varphi})
\alpha^{(N_f)}_r(\phi_3,\ldots,\phi_{N_f},\theta)\;,
\label{eq431}
\ee
where the phases $\alpha^{(N_f)}_r$ are given by
\be
\alpha^{(N_f)}_r(\phi_3,\ldots,\phi_{N_f},\theta)=\sum_{j=1}^{N_f-r}
e^{-i(j+r/2)(\phi_3\ldots+\phi_{N_f}-\theta)}
\beta^{(N_f)}_{j,j+r}(\phi_3,\ldots,\phi_{N_f})
\label{eq432}
\ee
with
\be
\beta^{(N_f)}_{k,l}(\phi_3,\ldots,\phi_{N_f})=
\left|\begin{array}{ccccccccc}
1 & \cdots & e^{i(k-2)\phi_3} & e^{ik\phi_3} & \cdots 
& e^{i(l-2)\phi_3} & e^{il\phi_3} & \cdots & e^{i(N_f-1)\phi_3} \\
\vdots & & \vdots & \vdots & & \vdots & \vdots & & \vdots  \\
1 & \cdots & e^{i(k-2)\phi_{N_f}} & e^{ik\phi_{N_f}} & \cdots 
& e^{i(l-2)\phi_{N_f}} & e^{il\phi_{N_f}} & \cdots & e^{i(N_f-1)\phi_{N_f}}
\end{array}\right|\;.
\label{eq433}
\ee
The integration over the angle $\varphi$ can be easily performed,
since
\bea
\label{eq434}
\int\limits_0^{2\pi}\frac{d\varphi}{2\pi}e^{\mu\psi\cos\varphi}
\left(e^{ir\varphi}-e^{-ir\varphi}\right)
& = & I_{r+1}(\mu\psi)-I_{r-1}(\mu\psi) \\ \nonumber
& = & -\frac{2r}{\mu\psi}I_r(\mu\psi)\;.
\eea
From that the final expression for the partition function with
degenerate masses is
\be
{\cal Z}^{(N_f)}(\theta,\mu)=-2e^{-\theta/2}\int\limits_0^{2\pi} 
\prod_{m=3}^{N_f}\left(\frac{d\phi_m}{2\pi}
e^{\mu\cos(\phi_m)-i(m-3/2)\phi_j}\right) \sum_{r=1}^{N_f-1}
r(-1)^r\frac{I_r(\mu\psi)}{\mu\psi}
\alpha^{(N_f)}_r(\phi_3,\ldots,\phi_{N_f},\theta)\;.
\label{eq435}
\ee

Alternatively, from eq.\ (\ref{eq35}), the equal mass partition function is
\begin{mathletters}
\bea
{\cal Z}^{(N_f)}(\theta,\mu) &=& \varepsilon_{a_1\ldots a_{N_f}}
        \sum_{\nu = -\infty}^{\infty} e^{i\nu \theta}
        \prod\limits_{m=1}^{N_f} I_{\nu+m-a_m}(\mu) \\
        &=& \varepsilon_{a_1\ldots a_{N_f}}
        \sum_{\nu = -\infty}^{\infty}e^{i\nu \theta}
        \prod\limits_{m=1}^{N_f}
        \int\limits_0^{2\pi}\frac{d\phi_{m}}{2\pi}
        \exp\left[\mu \cos(\phi_{m}) + i (\nu + m - a_m)\phi_{m}
        \right] \,\, .
\eea
\end{mathletters}
Using eq.\ (\ref{eq426}), the summation over $\nu$ gives 
a delta function which facilitates the integration $\phi_{a_1}$:
\bea
{\cal Z}^{(N_f)}(\theta,\mu) &=& 
        \frac{\varepsilon_{a_1\ldots a_{N_f}}}{(2\pi)^{N_f-1}}
        \int\limits_0^{2\pi} d\phi_{2}\ldots d\phi_{N_f}
        \exp\left[\mu \cos(\theta+\phi_2+\phi_3+\ldots+\phi_{N_f})\right]
        \\ \nonumber
        &\times&\exp\left[\mu \cos(\phi_{2})+\mu \cos(\phi_3) + 
        \ldots \mu \cos(\phi_{N_f}) \right]
        \\ \nonumber
        &\times&\cos\left\{(a_1-1)(\theta+\phi_2+\ldots+\phi_{N_f})
        + (2-a_2)\phi_2 + \ldots + (N_f-a_{N_f})\phi_{N_f}
        \right\} \,\, .
\eea
The integration over $\phi_2$ can be done analytically \cite{GR} 
by expanding 
$\cos(\theta+\phi_2+\ldots+\phi_{N_f}) = 
\cos(\phi_2) \cos(\theta+\phi_3+\ldots+\phi_{N_f})
-\sin(\phi_2) \sin(\theta+\phi_3+\ldots+\phi_{N_f})$.

The result is 
\bea
{\cal Z}^{(N_f)}(\theta,\mu) &=& 
	\varepsilon_{a_1\ldots a_{N_f}} \prod\limits_{m=3}^{N_f} 
        \left( \int\limits_0^{2\pi} \frac{d\phi_m}{2\pi}
        e^{\mu \cos(\phi_m)}\right) I_{a_1-a_2+1}\left(2 \mu \cos\left(
	\frac{\theta+\phi_3
        +\ldots+\phi_{N_f}}{2}\right)\right)\\ \nonumber
        &\times&
        \cos\left\{(3-a_1-a_2)(\theta+\phi_3+\ldots
	+\phi_{N_f})/2-(3-a_3)\phi_3-\ldots-
	(N_f-a_{N_f}) \phi_{N_f}\right\}\,\, .
\eea
Using the addition rules for modified Bessel functions, one arrives 
at eq.\ (\ref{eq:geneqZ}).

\end{document}